\documentclass{IEEEphot}

\usepackage{ulem}       
\usepackage{url}
\usepackage{tikz}
\usepackage{pgf}

\usepackage{amsmath}
\usepackage{amssymb}
\usepackage{graphicx}
\usepackage{subcaption}
\usepackage{url,hyperref, nameref}
\usepackage{tikz}
\usetikzlibrary{arrows,automata}

\usepackage{multirow,bigdelim,colortbl}

\usepackage{algorithmic}
\usepackage{algorithm}

\begin{document}

\title{Reliable and Efficient Broadcast Routing\\ Using Multipoint Relays Over VANET\\ For Vehicle Platooning}

\author{Xing Wang, Alvin Lim}
\affil{Department of Computer Sciences and Software Engineering\\ Auburn University, AL 36849 USA}  



\maketitle



\begin{abstract}
In this paper, we design and implement a reliable broadcast algorithm over a VANET for supporting multi-hop forwarding of vehicle sensor and control packets that will enable vehicles to platoon with each other in order to form a road train behind the lead truck. 
In particular, we use \textit{multipoint relays} (MPRs) for packet transmission, which leads to more efficient communication in a VANET. We evaluate the performance based on simulation by running a platooning simulation application program, and show that with MPRs, the communication in the VANET to form a road train is more efficient and reliable.
\end{abstract}



\section{Introduction}
\subsection{Background} \label{sec:bg}
A mobile \textit{ad hoc} network (MANET) is the cooperative wireless network formed by a collection of mobile hosts (MHs) without the support from fixed infrastructure or centralized access point. 
As a special subcategory of MANET, the vehicular \textit{ad hoc} network (VANET) is built up of moving vehicles in which each participating vehicle serves as a wireless router or node, and in turn creates a network with a wide range \cite{shrimali2017}. Due to the increasing demand of road safety and convenience, VANET is emerging rapidly and has many applications, such as vehicle collision warning, security distance warning, driver assistance, cooperative driving, cooperative cruise control, dessemination of road information, internet access, map location, automatic parking, driverless vehicles, etc \cite{paul2012vanet}.  In this project, we design and implement a VANET for self-driving vehicles using a \textit{cooperative adaptive cruise control} (CACC) platooning application.

\begin{figure}
    \centering
    \includegraphics[width=\textwidth]{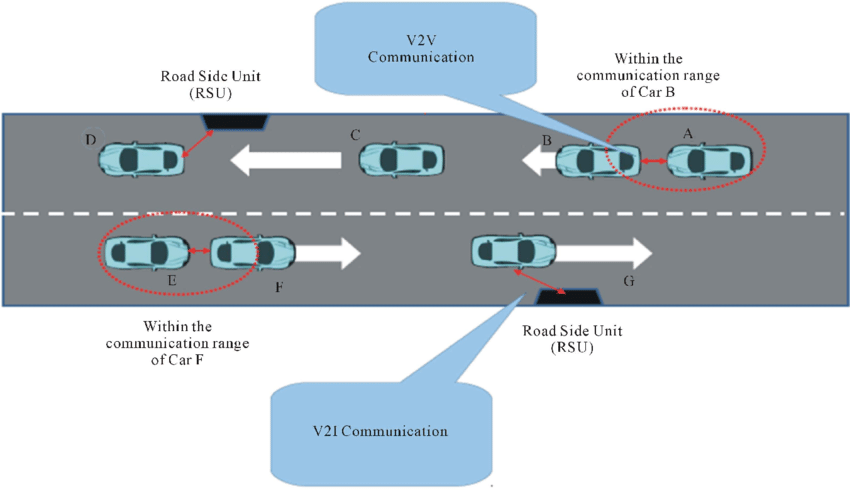}
    \caption{Vehicle-to-vehicle (V2V) and vehicle-to-infrastructure (V2I) communication in VANET, adapted from \cite{sari2015review} }
    \label{fig:v2vv2i}
\end{figure}

As compared to other MANETs, VANET has some unique characteristics, some are attractive while others make it challenging in real applications. For example, in other MANETs, the energy constraint often needs to be one of the most important design optimization criteria since many \textit{ad hoc} devices rely on batteries or other limited power source. In contrast, this is not an issue in VANETs as modern vehicles could afford powerful computing, communication and sensing capabilities. 
In most MANETs, nodes are free to move in arbitrary direction and thus the network topology is highly unpredictable. On the contrary, vehicles are allowed to move within only one or two directions on the same road, moreover, road and traffic information is usually available from positioning systems and map-based technologies (e.g., GPS), all make predicting vehicles' movements possible, and leave room for modeling the vehicles' behavior and further optimizing routing through data. However, VANETs have their own characteristics of extremely dynamic topology. 
Vehicular network will be frequently partitioned and dynamic nature of traffic may result in large inter vehicle gaps in sparse network.  
The vehicles move at high speed and change their position constantly in the dynamic scenarios. Network topology changes frequently as the link between the nodes connect and disconnect very often, and the frequent disconnection occurs more often in sparse network. 
We also note that in most VANETs, roadside infrastructure is also a vital component, 
and both types of communication, vehicle-to-vehicle (V2V) and vehicle-to-infrastructure (V2I), are significant requirements for building an advanced intelligent transportation system (ITS). Figure 
\ref{fig:v2vv2i} shows the two types of communication in VANET, in which the roadside infrastructure assists to create a better share of information between vehicles. In this project, however, we will not consider the V2I component for simplicity purpose, and focus on the V2V communication. 

The introduction of automation into vehicles in a connected vehicular environment is potentially transformative \cite{dey2016review}. To facilitate the research and implementation of automation in vehicles, a classification system based on six different levels (ranging from fully manual to fully automation) was published in 2014 by Society of Automotive Engineers (SAE) International, an automotive standardization body, as ``\textit{J3016, Taxonomy and Definitions for Terms Related to On-Road Motor Vehicle Automated Driving Systems}'', which was then updated in 2016, called ``\textit{J3016\_201609}'' \cite{sae}. The first automation driving support system, the conventional cruise control (CCC), allows drivers to drive at a certain speed. In recent years, with the improvement of vehicle control technology, the adaptive cruise control (ACC) system allows a vehicle to drive behind a leader at a certain distance, which improves roadway capacity and traffic safety
as well as fuel efficiency \cite{vahidi2003research}. This system is designed to further improve the driving experience such as comfort. Enabling the connectivity between vehicles and roadside infrastructure, ACC application could be extended to form a platoon known as cooperative adaptive cruise control (CACC) \cite{van2006impact} and belongs to level 2 automation defined by SAE. With the shared information between vehicles, the CACC allow vehicles in a platoon to maintain smaller headway compared to ACC \cite{xu2003simulation}. We will discuss the platooning system in Section 2.


\subsection{Objective and Research Questions}
The objective of this project is to design and implement a reliable broadcast algorithm over a VANET for supporting multi-hop forwarding of vehicle sensor and control packets for a CACC system that will enable vehicles to platoon with each other in order to form a road train behind the lead truck. 
In particular, we aim to use \textit{multipoint relays} (MPRs) for packet transmission. We expect using MPRs would lead to more efficient communication in a VANET. For comparison purpose, a \textit{reliable broadcast algorithm} needs to be implemented as well. We will evaluate their performance based on simulation by running a platooning simulation application program. 

To better understand the goal, we will first investigate what is known today about how to obtain better communication in platooning, this will be achieved by studying the existing research literature to address the following research questions:
\begin{itemize}
    \item What characterizes existing literature on platooning?
    \item What are the communication methods for platooning in literature?
    \item Which routing methods have been used in literature?
    \item What solution have been proposed in literature to improve communication? 
    \item Is MPR transmission better than regular flooding in a vehicle platoon? 
\end{itemize}

\subsection{Outline}
This paper is organized as follows: 
Section 2 presents an overview of vehicle platooning, which serves as the background of our project. 
Section 3 discusses routing protocols as applied to VANETs, especially those relevant to the implementation in this project.
Section 4 describes the design of the models. 
In Section 5, we provide the details for implementation and environment setup. Section 6 shows the experimental results in our simulation. 
And finally, we conclude in Section 7.

\section{Vehicle Platooning}

Vehicle platooning can be illustrated as a chain of vehicles travelling together at a given speed and distance. It is a concept in the automotive industry that mainly aims at increasing the road capacity and improve the travel efficiency as well as safety. Under cooperative driving, automated vehicles can be organized in tightly controlled groups, also called \textit{platoons} (or \textit{road trains}). Throughout this report, we will use \textit{platoon} and \textit{road train} interchangeably.

\begin{figure}
    \centering
    \includegraphics[width=.9\textwidth]{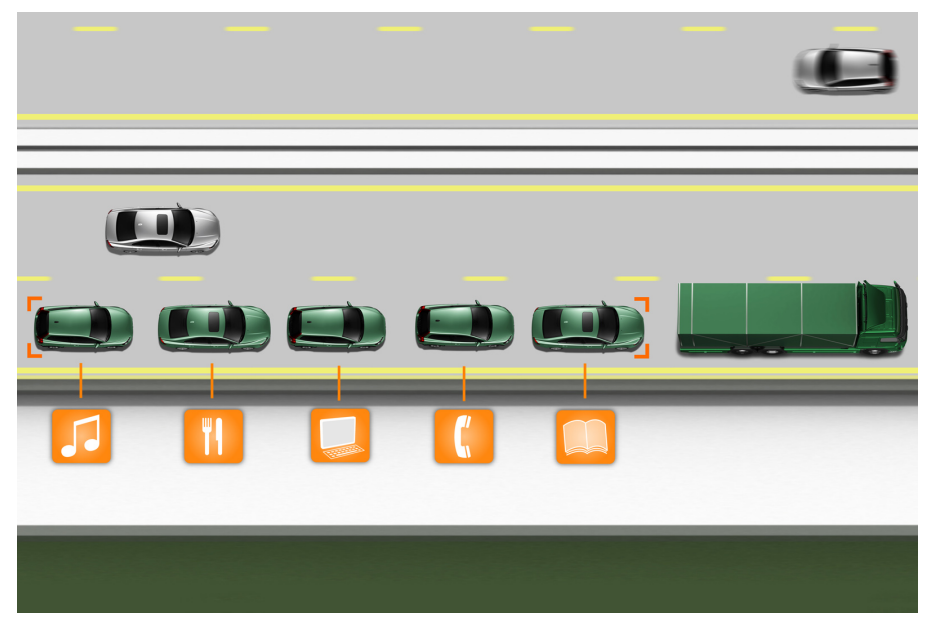}
    \caption{An illustration of vehicle platooning, adapted from \cite{robinson2010operating}}
    \label{fig:platoon}
\end{figure}

Figure \ref{fig:platoon} shows an illustration of vehicle platoon as a SARTRE demo (see Section \ref{sec:deploy}), in which several vehicles travel along the highway as a single unit. These vehicles follow one another with a small spacing and are ``linked'' through some communication and control mechanism. The first vehicle in the platoon, the lead truck, continuously provides the others, the following vehicles, with necessary information upon which the platooning could be executed continuing.

\subsection{Overview of Platooning Deployments} 
The idea of automated highway traffic was presented by General Motors way back to 1940 World's Fair in New York, and the early implementation, in which vehicle platooning plays an important role, can be traced back to 1980s in California \cite{kavathekar2011vehicle}. 
Ever since then, many projects and experiments have shown that platooning is feasible in practice. Here we provide a brief overview of several famous platooning projects over the world, mainly cited from \cite{bergenhem2012overview} as well as the corresponding project reports.

\subsubsection*{SARTRE} \label{sec:deploy}
Safe Road Trains for the Environment (SARTRE)  is a European Commission Co-Funded FP7 project to investigate and trial technologies and strategies for the safe platooning of road vehicles. 
The three-year project was launched in 2009. The research and development was carried out by several European auto manufactures with Volvo at the lead. A first practical test successfully took place in December 2010. \cite{robinson2010operating}

The SARTRE definition of platooning implies that the platoon is led by a vehicle which is manually driven by a professional driver, such as a truck. This driver must have a valid licence and is assumed to have additional training for leading a platoon.
The following vehicles (trucks and passenger cars) are under automated longitudinal and lateral control of the lead vehicle, so than these drivers are hence able to undertake other tasks such as using mobile phones. 
The following vehicles may join or leave the platoon dynamically, e.g. leave on arrival at the desired destination. In SARTRE, a car can only be a following vehicle in a platoon. 

The project aims to explore technology for platooning on roads without changes to the infrastructure and that it is safe enough to allow mixing with other users of public roads. Expected advantages of platooning include a reduction in fuel consumption, increased safety, traffic efficiency, increased driver convenience and comfort. \cite{bergenhem2010challenges}

One of the key characteristics of SARTRE is that its platooning application requires that V2V communication is used in addition to local sensors in each vehicle. With V2V implies that data can be sent directly from the source rather being indirectly measured locally with sensors. 
Using only local sensors can lead to lateral and longitudinal instability, increasing oscillations, and unsafe behaviour of the platoon.
V2V communication allows sharing of local vehicle signals such as speed and sensor data among vehicles in the platoon. The shared signals are used in the control algorithms of the platoon. 
\cite{bergenhem2012vehicle}

In January 2011, SARTRE made the first successful demonstration of its platooning technology at the Volvo Proving Ground near Gothenburg, Sweden, in which a lead truck was followed by a car. In January 2012, SARTRE made a second demonstration in Barcelona, Spain, in which a lead truck was followed by three cars driven entirely autonomously at speeds of up to 90 km/h (56 mph) with a gap between of at most 6 m (20 ft). 

\subsubsection*{PATH}
The Partners for Advanced Transit and Highways (PATH) program in California has been involved in the research of connected and autonomous vehicles for over 30 years. 
The PATH research on automated platoons was initially motivated by the need to produce a significant increase in the capacity of a highway lane, so that increases in travel demand could be accommodated with a minimum of new infrastructure construction. The PATH kinematic studies of highway capacity showed that it could be possible to increase passenger car lane capacity by a factor of two to three over today’s capacity if the vehicles were driven in
platoons of up to ten cars [7]. The gaps between platoons would be long enough to ensure that even in the worst crash hazard condition, with maximum deceleration, a following platoon would be able to stop without hitting the last vehicle of the forward platoon, even while there could be low-speed crashes between vehicles within the forward platoon. An extensive modelling and simulation study of crash safety and capacity showed the advantages of the platoon mode over individual automated vehicles [8]. The PATH studies have been based on the assumption that all vehicles would be automated, including the first vehicle of the platoon, in order to maximize efficiency and remove the potential for driver errors to cause crashes.

PATH first tested the longitudinal control of a 4-car platoon at 4 m separation at highway speeds in 1994, and then developed the 8-car automated platoon for the National Automated Highway System Consortium (NAHSC) public demonstration in August, 1997. Over 1000 visitors were given passenger rides in these platoon vehicles, which did a variety of manoeuvres including lane changing and joining and leaving the platoon as well as normal car following, all under completely automatic control. All of the intelligence, providing for sensor signal processing, V2V communication and coordination, and lateral and longitudinal control, was done on a single Pentium computer running at a 166 MHz rate, so this did not require high computational power. The gaps between the vehicles within the platoon were maintained with a 20 cm RMS error, which is small enough that vehicle occupants felt as if they had a mechanical coupling to the preceding car, while also maintaining a smooth ride quality for
comfort \cite{rajamani2000demonstration}. 

Another focus of the PATH platooning was on heavy trucks, mainly because of the potential for energy saving. Forming trucks in automated platoons of 3 individuals could enable a capacity of about 1500 trucks per lane per hour, which is twice the capacity achievable with trucks driven individually. The PATH experiments on truck platoons have shown the technical feasibility of driving two trucks at a gap of 3 m and three trucks at a gap of 4 m between trucks. These experiments also showed direct fuel consumption savings of about 5\% for the lead truck and a range of 10\% to 15\% for the following trucks under conservative estimations.
\cite{browand2004fuel}

\subsubsection*{Energy ITS}
\begin{figure}
    \centering
    \begin{subfigure}[b]{0.475\textwidth}
        \includegraphics[width=\textwidth]{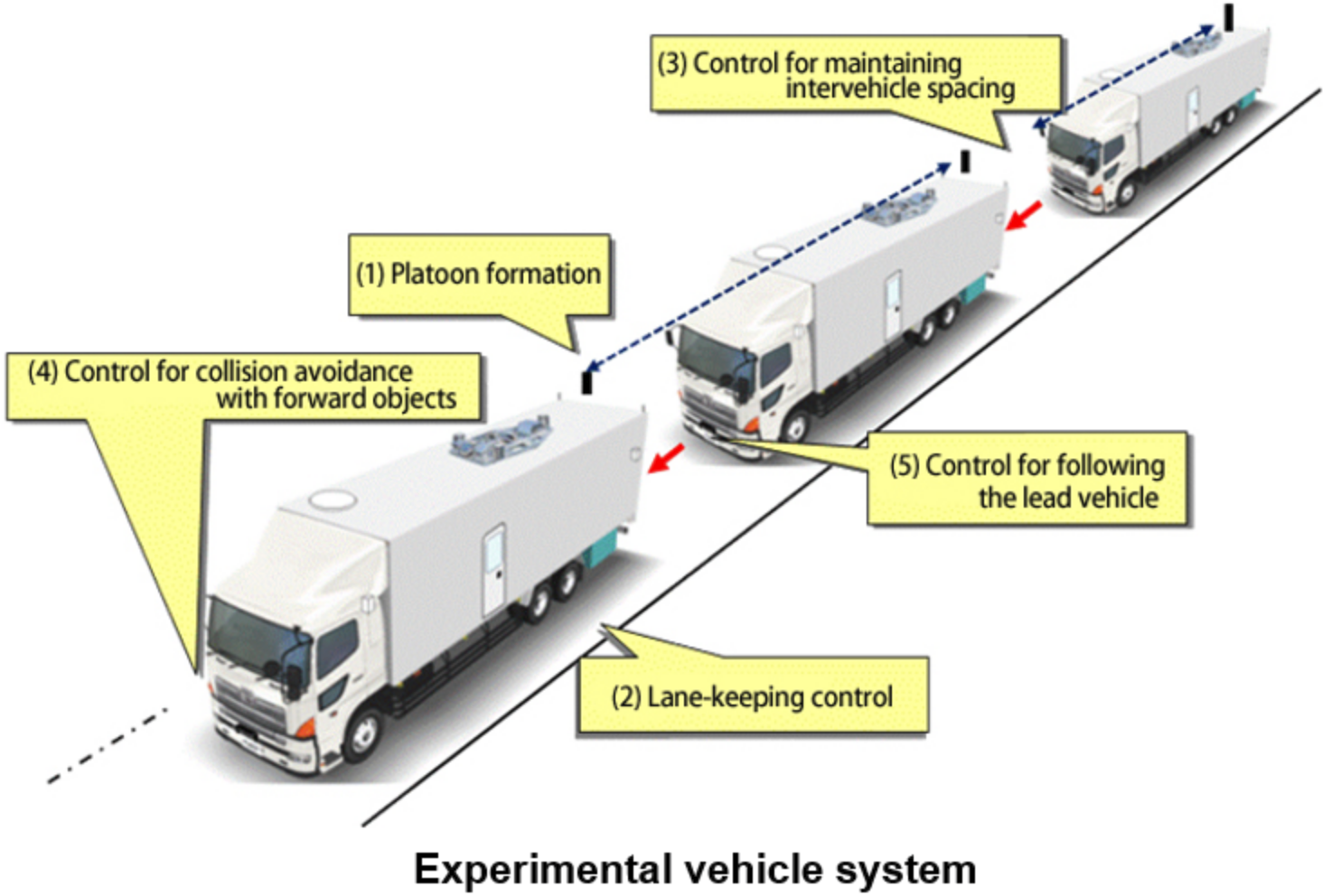}
    \end{subfigure}
    ~ 
    \begin{subfigure}[b]{0.475\textwidth}
        \includegraphics[width=\textwidth]{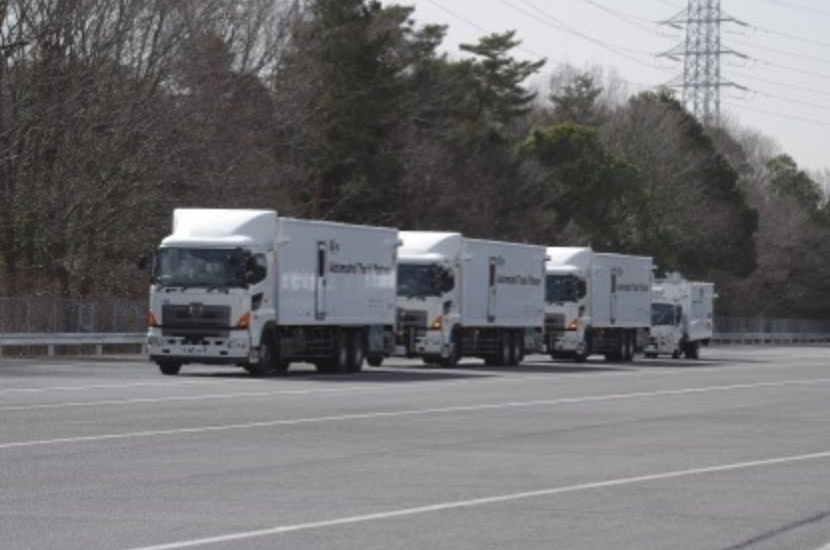}
    \end{subfigure}
    \caption{Vehicle platoon demo of 2008-2012 Energy ITS (Japan). Source: \url{http://itsasia-pacific.com/about-its-asia-pacific/examples-of-its-deployment-by-countryarea/2008-2012-energy-its\%EF\%BC\%88japan/} }
    \label{fig:japan}
\end{figure}

2008-2012 Energy ITS (Japan) was a Japanese national project for International Task Force (ITS) on Vehicle Highway Automation, funded by Japanese Ministry of Economy, Trade and Industry. It aimed at energy saving and global warming prevention with ITS technologies, and has two themes: an automated truck platoon and an evaluation method of effectiveness of ITS on energy saving. 
The lateral control is based on the lane marker detection by the computer vision, and the longitudinal control is based on the gap measurement by radar and lidar in addition to the V2V communications \cite{tsugawa2013overview}. See Figure \ref{fig:japan}.

The project succeeded in developing automated truck platooning technologies for operating four trucks with an inter-vehicle spacing of 4 meters and traveling at a speed of 80 km/h, and achieved to reduce energy by about 15 \% (measurement) by the aerodyamic drag reduction, and CO$_2$ by 2.1 \% along an expressway (simulation) when the 40 \% penetration in heavy trucks by the roadway capacity increase \cite{tsugawa2011automated}. 

\subsubsection*{Others}
The Federal Highway Administration in 2013 funded two research projects in heavy truck platooning (without steering automation). One is led by Auburn University, while the other is led by California Department of Transportation, with UC Berkeley and Volvo Trucks.

\subsection{Benefits of Vehicle Platooning}
As we mentioned as the main goals for each of the famous deployment, vehicle platooning offers a number of benefits. First of all, most platooning projects aim towards the reduction of fuel consumption for Heavy Duty Vehicles (HDV). 
HDV platooning for emission reduction and energy efficiency is intensively studied. It is known that driving at a short inter-vehicle spacing to a vehicle ahead results in a reduced fuel consumption and drivers are doing so today with increased stress levels. 
Since up to 1/4 of the fuel consumption for a typical HDV can be spent on overcoming the aerodynamic drag, forming in a platoon can significantly lower the air drag. Even the leading vehicle takes advantage of platooning at short inter-vehicle distances, due to reduced adverse aerodynamic effects.
An illustration of this effect is given in \cite{bonnet2000fuel}, which shows that trailing HDVs traveling at 80 km/h experience a 21\% fuel reduction when the distance between the vehicles is 10 m, while the fuel reduction is 16\% for an intervehicle distance of 16 m. The fuel reductions for the same vehicles and distances traveling at 60 km/h are approximately 16\% and 10\%, respectively. 
Therefore, it is clear that platooning has the potential to provide substantial economic benefits for individual haulage companies. 
Moreover, since vehicles account for a large proportion of total carbon emissions (20\% according to \cite{delft2012marginal}, and 1/4 comes from HDVs), lower fuel consumption can substantially make progress toward carbon reduction goals and achieve clear environmental gains.

Apart from reducing fuel consumption, platooning potentially offers other important gains. 
As vehicles are operated at close distances, the existing road infrastructure can be exploited more effectively and capacity can be increased. For passenger cars, it is argued that a capacity increase of 200\% can be achieved \cite{alam2015heavy}. In addition, platooning can provide a major improvement in the utilization of existing road capacity by significantly reducing road congestion.

Because human drivers are generally not capable of safely maintaining the close distances as required by platooning, it is clear that automated driving technologies are needed. The increased level of automation provides a second advantage of cooperative transport technologies since automation is generally believed to have a positive effect on road safety \cite{alam2016intelligent}. Namely, automatic systems can usually react more quickly to dangerous situations and can exploit additional information resulting from the communication and cooperation between vehicles. 
Moreover, the majority of traffic accidents are caused by human errors. Speeding, distractions, fatigue and even drunk driving are some of the reasons. In fact, it was reported that the drivers are partly or fully responsible for over 90\% of the traffic accidents \cite{hobert2012study}.
And increased automation has a strong potential to mitigate this factor. Moreover, autonomous driving system and vehicle platooning can greatly improve driver's comfort, in the way that the drivers may delegate certain driving tasks, allowing the driver to shift his/her focus, this can relief stress and especially fatigue for long trips. This in turn can improve the driving safety.

We summarize the benefits of vehicle platooning we discussed above:
\begin{itemize}
    \item Improvement of energy efficiency and transportation profitability
    \item Reduction of environmental impact
    \item Improvement of road capacity
    \item Improvement of safety and comfort, etc.
\end{itemize}

\subsection{Architecture of Vehicle Platooning System}
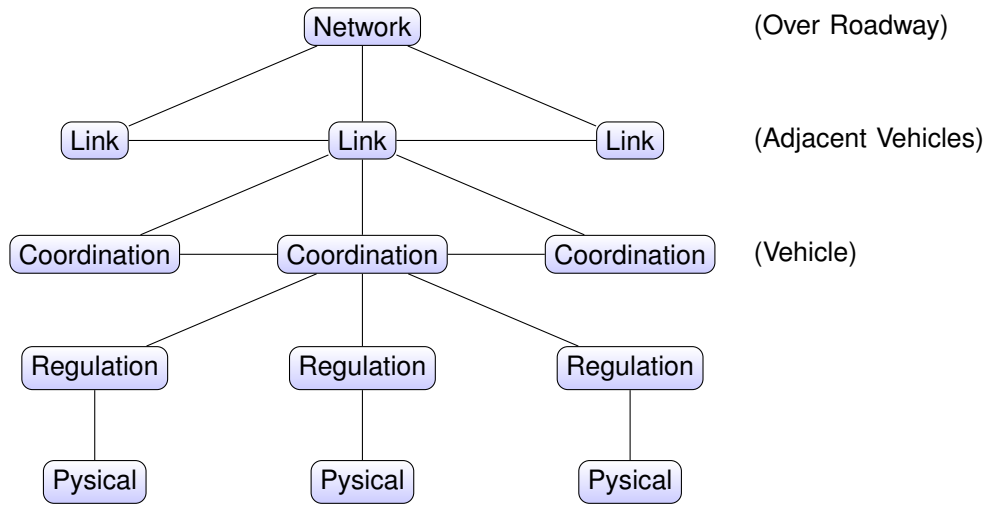
\begin{figure}
    \centering
    \begin{tikzpicture}[sibling distance=10em,
    	every node/.style={shape=rectangle, rounded corners, draw, align=center,
    				top color=white, bottom color=blue!20}
    	]
    	\node (n0) {Network}
            child {node (l1){Link}}
            child {node (l2){Link}
                child {node (c1){Coordination} }
                child {node (c2){Coordination} 
                    child {node {Regulation}    child {node {Pysical} }  }
                    child {node {Regulation}    child {node {Pysical} }  }
                    child {node {Regulation}    child {node {Pysical} }  }
                }
                child {node (c3){Coordination} }
            }
            child {node (l3){Link}}
        ;
        \path[draw,-] (l1) edge (l2); 
        \path[draw,-] (l2) edge (l3); 
        \path[draw,-] (c1) edge (c2); 
        \path[draw,-] (c2) edge (c3); 
        \begin{scope}[every node/.style={right}]
            \path (n0 -| c3) ++(15mm, 0) node {(Over Roadway)} ;
            \path (l3 -| c3) ++(15mm, 0) node {(Adjacent Vehicles)} ;
            \path (c3 -| c3) ++(15mm, 0) node {(Vehicle)} ;
        \end{scope}
    \end{tikzpicture}
    \caption{Hierarchical architecture for vehicle platooning}
    \label{fig:hierarchical}
\end{figure}

This section mainly cites \cite{shladover2005automated}.
A simple hierarchical framework for decomposing the implementation of a vehicle platooning system (or indeed any intelligent transportation system (ITS)), as shown in Figure \ref{fig:hierarchical}, has been established and found to be very useful for decades of research and development work, however, it is worthwhile to note that there is a significant spreading from each layer to the next. For instance, a large urban region might have a single network layer, but potentially hundreds of links. Each link could have dozens of clusters of closely coordinated vehicles, each of which could have a dozen or more vehicles \cite{shladover2005automated}.

The network layer maintains desired cruising speed throughout the network (in the absence of incidents or failures) by metering access to the network. 
The link layer takes care of issues such as managing flows around incidents, balancing traffic across lanes,  metering the entry rate of vehicles to prevent upstream propagation of congestion shock waves, and assigning suitable speed limits to each lane. 
Starting from the coordination layer, the emphasis shifts towards individual vehicles and a single platoon of vehicles rather than the aggregate traffic flows. 
At the coordination layer, vehicle trajectories are planned, information about traffic and road conditions is exchanged among vehicles, and manoeuvres are coordinated among adjacent vehicles, such as lane changing, joining and leaving a platoon, etc. 
The regulation layer is responsible for the control of vehicle motions, such as steering, acceleration, braking and spacing relative to the preceding vehicles, often the classic closed-loop control is applied. 
Finally, the physical layer is where in-vehicle sensing of position relative to the lane and other vehicles, the actuation of steering, engine, brakes, are performed. Most of the components used at the physical layer were adopted from development of sensor, computer vision, and actuator technologies. 

In this project, our design and implementation will not consider managing the traffic flow, but only pay attention to a single platoon, thus the network layer and link layer are not our concern. Nor shall we focus on the lower layer physics and control, but in order to better understand the vehicle behaviors in a platoon, the platooning control functionality is briefly discussed in Section \ref{sec:ctrl}.

\subsection{Platooning Control Functionality} \label{sec:ctrl}
The main regulation layer functions are lateral and longitudinal control of vehicle motions, as well as maneuver coordination of vehicle behaviors in platooning. We discuss these functionality in more detail below.

\subsubsection{Lateral Control}
The primary functionality of lateral control is keeping the vehicle following the lane center, typically requires solving a closed-loop regulator problem to steer the vehicle. Lateral control is in addition concerned with lane changing, when the vehicle needs to switch from following the center of one lane to following the center of the next lane.

\paragraph{Lane Tracking}
Lane tracking is the functionality that keeps a vehicle centered in a lane. It is a regulator control problem, in which the primary design considerations involve the fundamental trade-off between tracking accuracy and ride quality. Higher tracking accuracy enhances driving safety and makes it possible to save roadway construction cost by reducing the lane width, but it can also lead to harsher, jerkier lateral vehicle motions.  Challenges to lane tracking include difficult vehicle dynamics, and particularly non-linear speed-dependent response characteristics.

For steering a vehicle towards the center of a lane, it is important to know the current lateral position of the vehicle. Several technologies have been considered to measure this position, such as permanent-magnet tracking, computer vision and Differential Global Positioning System (DGPS), an enhancement of GPS that provides a higher accurate location service.

\paragraph{Lane Changing}
Lane changing is the functionality to steer a vehicle from the current lane to an adjacent lane. 
Changing lanes is considerably more challenging than simply following a lane, since it involves more complicated vehicle dynamic behaviour as well as the change of front target and coordination between vehicles.

In order to support this task, a magnet tracking system can be used.
Magnet-based lane changing can be done either by following a trail of magnets between the lanes at fixed locations, similar to
railroad switch that guides a train to an adjacent track (``infrastructure guided''); or by dead reckoning between the magnet trails in the center of the two lanes (``free lane change''), which involves using a lateral displacement estimator, based on acceleration and yaw rate measurements, combined with trajectory planning. 
And recently, the magnet infrastructure requirements is generally removed due to the progress of computer vision in recognizing the lanes.

\subsubsection{Longitude Control}
Longitudinal control of a road vehicle means controlling its speed and its separation from the preceding vehicle, requiring activation of engine, brake and possibly transmission actuators on the vehicle.
Analogous to lateral control, the primary requirements are tracking accuracy and ride comfort, which must be traded off against each other.
Ride comfort is important for driver and passenger acceptance but is also closely related to the ability of the control system to save fuel and to reduce emissions. While tracking accuracy is important for maintaining safety and for enabling vehicles to operate in close proximity to each other so that they can increase lane capacity and reduce aerodynamic
drag (thereby also saving fuel and emissions).

Longitudinal control implementations were typically based on use of radar sensors to measure vehicle-to-vehicle distance, but some systems have also used computer vision and more specialized image-processing sensor systems.
The differences among systems based on these alternative sensors, however, are not as significant as the differences between the autonomous vehicle following systems that rely only on their own sensors and that rely on cooperative control. 
In a cooperative vehicle following system, the controllers typically augment sensor data with wireless communication in order to actively collect state information (position, speed, acceleration and maneuver information) of other vehicles. These types of controllers are not only able to maintain string stability under a constant time gap, like autonomous controllers, but also under constant distance gaps. As these constant distance gaps are independent of the operating speed of the vehicles, the cooperative controllers are capable of keeping the headway small, even at high speeds. This is one of the reasons why cooperative controllers, as used in CACC systems, can achieve a much higher road capacity.

\subsubsection{Maneuver Coordination}
Besides longitudinal and lateral control, vehicle platooning involves the coordination of vehicle maneuvers. These maneuvers are typically the formation and splitting of platoons, the merging of traffic streams and coordination of changing lanes. This research mainly concerned with the joining and leaving a platoon. In the following sections we limit our description to these maneuvers.

\begin{figure}[htb]
    \centering
    \begin{subfigure}[b]{0.8\textwidth}
        \includegraphics[width=\textwidth]{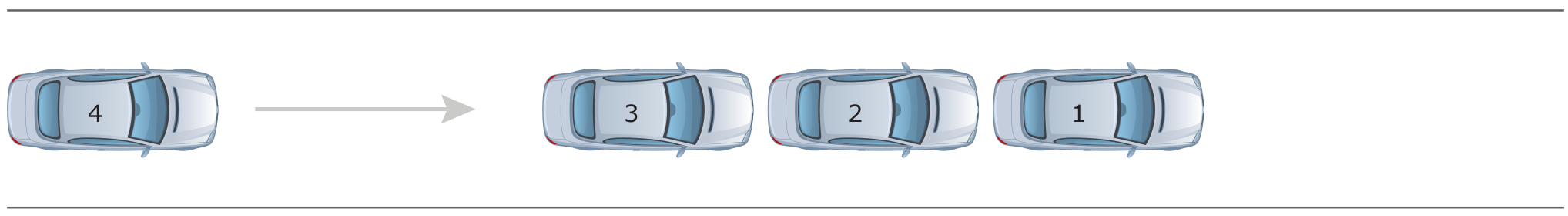}
        \caption{Before Joining}
    \end{subfigure}
    \\ \vspace{.1in}
    \begin{subfigure}[b]{0.8\textwidth}
        \includegraphics[width=\textwidth]{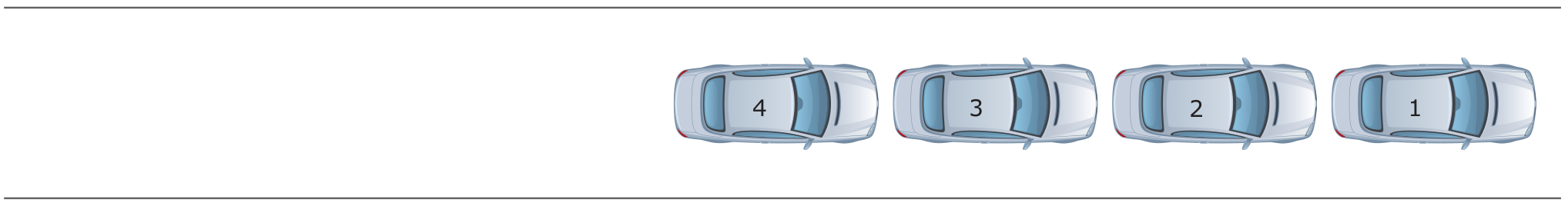}
        \caption{After Joining}
    \end{subfigure}
    \caption{Joining platoon}
    \label{fig:join0}
\end{figure}

\paragraph{Joining Platoon}
\textit{Joining} is the term used when a vehicle (or more) is combined in a platoon. This situation can be easily extended to the situation where two platoons are combined into a single one in general. Figure \ref{fig:join0} shows an example where vehicle 4 accelerates and joins at the back of a platoon (i.e., vehicles 1 to 3).  
We note that this is not the only way of joining a platoon: a vehicle may also join the platoon in the middle or in front of a platoon.

\begin{figure}[htb]
    \centering
    \begin{subfigure}[b]{0.8\textwidth}
        \includegraphics[width=\textwidth]{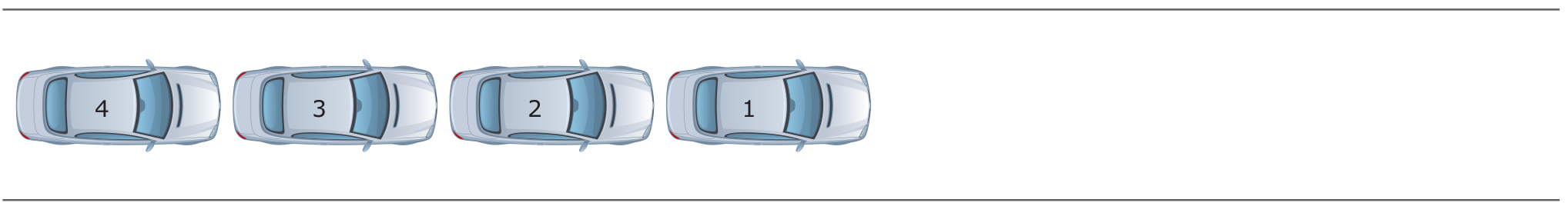}
        \caption{Before Leaving}
    \end{subfigure}
    \\ \vspace{.1in}
    \begin{subfigure}[b]{0.8\textwidth}
        \includegraphics[width=\textwidth]{figs/afterjoin.png}
        \caption{After Leaving}
    \end{subfigure}
    \caption{Leaving platoon}
    \label{fig:split0}
\end{figure}

\paragraph{Leaving Platoon}
A vehicle or a number of vehicles may leave the platoon and need not be bound by the rules for maintaining the platooning configuration. An example of this situation is illustrated in Figure \ref{fig:split0} where the last vehicle (i.e. vehicle 4) leaves the platoon. We also note that a vehicle in any position of the platoon may leave.

By coordinating the maneuvers among the involved vehicles, an improvement of the efficiency could be enabled. Such automatic coordination depends heavily on wireless communication. An introduction to wireless communication in vehicular environments is provided in the next section.

\subsection{Control Models}
\subsubsection{Vehicle Longitude Dynamics}
The following equation is widely employed to model the nonlinear longitudinal dynamics
\[ \label{eqn:nonlinear}
\begin{cases}
    \dot{x}_i(t) = v_i(t) \\
    m_i \dot{v}_i(t) = \dfrac{\eta_{T, i}}{r_i} - c^A_i v_i^2 - m_i gf \\
    \tau_i \dot{T}_i(t) + T_i(t) = T^{des}_i(t)
\end{cases} 
\]
in which $x_i(t), v_i(t)$ denote the (longitude) position and velocity of vehicle $i$, respectively. $m_i$ is the vehicle mass, and $r_i$ is its tire radius. $\eta_{T, i}$ denotes the mechanical efficiency of driveline, $c^A_i$ denotes its aerodynamic drag coefficient, $g$ is the gravitational acceleration, $f$ is the rolling resistance coefficient; $\tau_i$ is the inertial delay of vehicle longitudinal dynamics; $T_i(t)$ denotes the actual driving/braking torque, while the desired quantity is $T_i^{des}(t)$. 
The position and velocity of the leading vehicle are denoted by $x_0(t)$ and $v_0(t)$, respectively.

This nonlinear model is close to the real vehicle dynamics and some studies directly use it for platooning control. However, explicit performance limits are rather difficult to analyze with given spacing policy and communication topology. 
Rather, linear models are more frequently used for tractable issues. 

As the simplest case, the single integrator model which takes the vehicle speed as the input variable to be controlled, i.e., $\dot{x}_i(t) = u_i(t)$. While it significantly simplifies the theoretical analysis on controller design, however, the single integrator is unrealistic since controlling of the speed introduces a lot of instability. An improvement of that is to use the double integrator, i.e.,
\(
\begin{cases}
    \dot{x}_i(t) = v_i(t) \\
    \dot{v}_i(t) = u_i(t)
\end{cases} 
\)
in which the acceleration of the vehicle serves as the control variable $u_i(t)$. Many important theoretical results rely on the assumption of second-order dynamics which still does not catch many features of real vehicle dynamics, e.g., inertial delay in platoon dynamics \cite{li2017dynamical}. A common solution is to further add a state to approximate the input/output behaviors of the platoon dynamics, which equivalently settle the control input to engine/brake torque, etc. Most approximations use either feedback linearization, or lower-layer control technique, and the resulting model is often called the third-order model, which is the most frequently used one in CACC dynamics:
\(
\begin{cases}
    \dot{x}_i(t) = v_i(t) \\
    \dot{v}_i(t) = a_i(t) \\
    \dot{a}_i(t) = -\dfrac{1}{\eta_i} a_i + \dfrac{1}{\eta_i} u_i
\end{cases} 
\)
where $\eta_i$ is the internal actuator dynamics parameter. We can write it in the state space form:
\begin{align*}
    \dot{\mathbf{x}}_i(t) & = \mathbf{A}_i \mathbf{x}_i(t) + \mathbf{B}_i\mathbf{u}_i(t) \\
    \mathbf{x}_i(t) & = \begin{bmatrix} x_i \\ v_i \\ a_i \end{bmatrix}, \mathbf{A}_i = \begin{bmatrix} 0 & 1 & 0 \\ 0 & 0 & 1 \\ 0 & 0 & -\dfrac{1}{\eta_i} \end{bmatrix}, \mathbf{B}_i = \begin{bmatrix} 0 \\ 0 \\ \dfrac{1}{\eta_i} \end{bmatrix}    
\end{align*}

Alternatively, the corresponding transfer function can be written as
\begin{align*}
    X_i(s) & = \dfrac{1}{s}Y_i(s), \quad Y_i(s) = \dfrac{1}{s}Z_i(s) \\
    Z_i(s) & = H_i(s)U_i(s)  
\end{align*}
in which $X_i, Y_i, Z_i$ are the Laplace transforms of vehicle $i$'s position, velocity, and acceleration, respectively; $U_i$ is the control input, and $H_i$ is a linear Single-Input-Single-Output (SISO) transfer function which is strictly proper. Clearly, this model has two integrators and a lower order inertial delay, which leads to some fundamental limitations for certain platoons. 

In vehicle platooning, all vehicles in a platoon have the same objective to follow a leading vehicle with a certain distance, which is the acceptable/comfortable inter-vehicle distance determined by the spacing policy. There are two major spacing categories that adopted by existing car-following models \cite{dey2016review}. 
The first category has a fixed desired inter-vehicle distance \cite{peters2014leader}, while the second one is the so-called ``velocity dependent spacing'' category, which determines the inter-vehicle distance based on vehicle velocity. The latter is more commonly used, in particular, for a vehicle $i$ in the platoon, the velocity dependent spacing policy relates the desired inter-vehicle distance with the time headway as $$ d_{r, i} = r_i + h_{d, i}v_i$$
where $d_{r,i}$ is the desired inter-vehicle distance between vehicle $i$ and the one in front of it, $r_i$ is the standstill distance, $h_{d,i}$ is the headway-time constant, representing the time that it will take the vehicle to arrive at the same position as its predecessor when $r_i = 0$, and finally, $v_i$ is the velocity of vehicle $i$. The actual distance $d_i$ between two consecutive vehicles is then given by $$d_i = x_{i-1} - (x_i + L_i), $$
where $x_i$ is the position of vehicle $i$, and $L_i$ is its length. The local control objective, which is referred to as vehicle following, can now
be defined as regulating the spacing error to zero, where the error is defined as 
$$e_i = d_i - d_{r, i}. $$

The control of the collective behavior of a platoon is based on every vehicles' mutual awareness of their states (e.g., inter-vehicle distance, vehicle speed and acceleration, as we mentioned above), which is achieved by inter-vehicle sensing and communication. 
The information provided by inter-vehicle sensing and communication serves as an important input to the controllers in each vehicle, thus having a significant impact on the platooning behavior. 
In literature, the connectivity of information exchange between vehicles is captured by the so-called \textit{information flow topology}, and affects the platoon performance such as \textit{string stability}, \textit{stability margin}, and \textit{coherence behavior}. These performance measures can often be used as objectives or constraints to form the control as an optimization problem. 

\begin{figure}
    \centering
    \includegraphics[width=\textwidth]{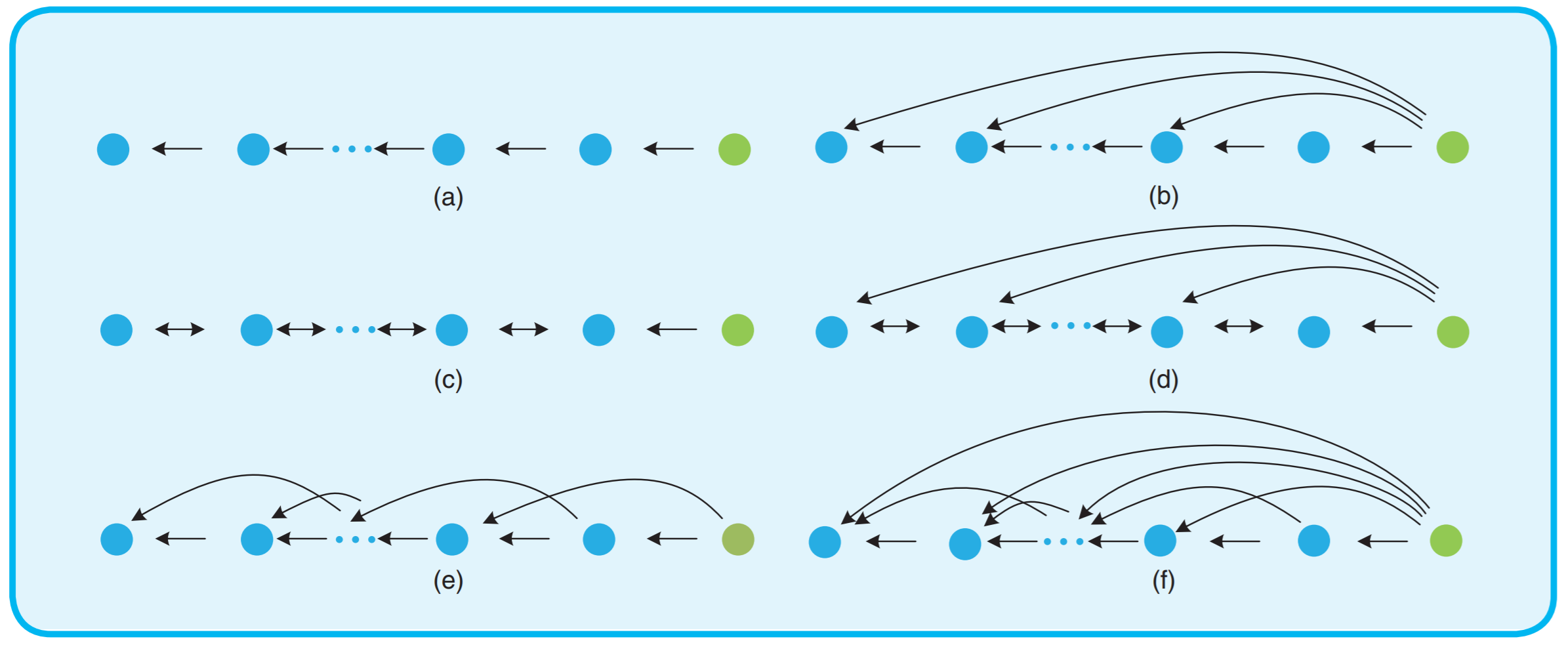}
    \caption{Common categories of information flow topology, (a) PF, (b) PFL, (c) BD, (d) BDL, (e) TPF, (f) TPFL. Source: \cite{li2017dynamical}, note that the original caption has wrong denotation.}
    \label{fig:ift}
\end{figure}

There are plenty of information flow topologies studied by researchers.
Early-stage platoon control is mainly based on radar-sensing, and a vehicle can only obtain information about its nearest neighbors, i.e., immediately preceding and following vehicles. Typical topology types include the Predecessor Following (PF) and Bi-Directional (BD) topologies, as shown in Figure \ref{fig:ift} (a) and (c), respectively. Nowadays, with V2V communication via technologies such as IEEE 802.11p-based DSRC and the emerging 5G solutions, various topologies become feasible since a vehicle can communicate with vehicles beyond its immediate surroundings. More types of topology have been derived, such as the Predecessor-Following Leader (PFL), Bi-Directional Leader (BDL), Two Predecessor-Following (TPF), and Two Predecessor-Following Leader (TPFL). See Figure \ref{fig:ift}. 
Since information flow topology has a significant impact on the behavior of a platoon, it is important to adapt the topology (e.g., by controlling the transmission power of communications) based on the need of platoon control.

\begin{figure}
    \centering
    \includegraphics[width=\textwidth]{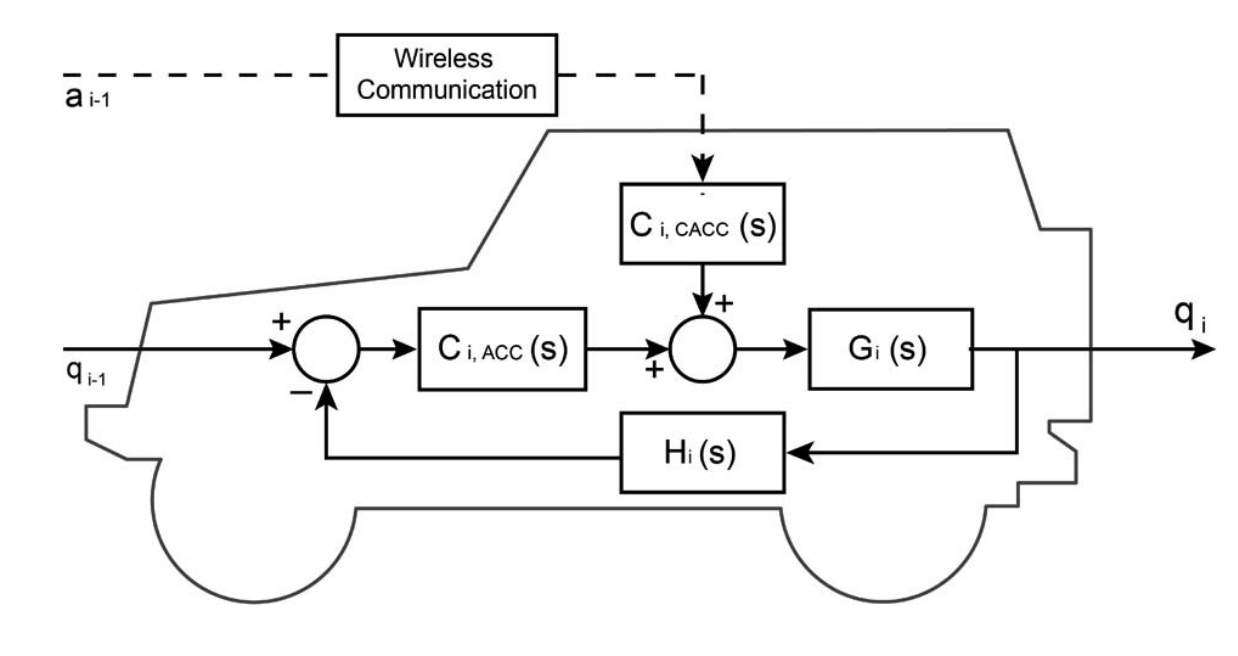}
    \caption{Block diagram showing the control structure of one vehicle in a CACC platoon with PF topology, apapted from \cite{li2017dynamical}}
    \label{fig:block}
\end{figure}

The most commonly used topology is predecessor following (PF) where the following vehicle only receives communication signal from its predecessor. 
As an example, Figure \ref{fig:block} shows a block diagram of a vehicle in a CACC platoon under this type, in which the controller can be seen as two parts: feedback ACC ($C_{i,ACC}$) and feed-forward CACC ($C_{i,CACC}$). 
Both the CACC and ACC inputs can be seen as acceleration signals that act as an input to the vehicle dynamics. The conventional ACC controller is usually a simple \textit{proportional derivative} (PD) or \textit{proportional integral derivative} (PID) controller. The controller is designed to process the wireless communication signal. 
A feed-forward filter is used to process the acceleration data that are transmitted from the preceding vehicle to obtain the control input. 
Under the predecessor-leader following topology, the controller takes two error signals (i.e., the spacing error with its predecessor and the spacing error with the lead vehicle) into account since the subject vehicle has communication with both its predecessor and lead vehicle in the platoon. PID controller is also applied in the controller design. 
Bidirectional topology allows the subject vehicle communicate with its adjacent vehicles, and the controller uses the front and rear spacing error as the feedback signals. 

The environment for platooning and autonomous driving technologies always has uncertainties. The control systems need to have the ability to deal with stochastic disturbance and risks. The stochasticity can have various sources, such as the drivers or the system itself. \cite{desjardins2011rl} proposed a reinforcement learning approach for the cooperative adaptive cruise control. The authros modeled the environment as a Markov Decision Process (MDP), and incorporated the stochastic game theory into the system to improve the CACC performance. Their simulation results show that small disturbances are damped through the platoon. 
And in \cite{van2011approximating}, the uncertainties in the communication network and sensor information were studied. The uncertainty parameters and delays were approximated to be Gaussian distributed, and this model was used to calculate the minimal time headway for safety. 

\subsection{Vehicular Wireless Communication}
There are several technologies that can support the wireless vehicle-to-vehicle (V2V) and vehicle-to-infrastructure (V2I) communication such as IEEE 802.11 (WiFi), IEEE 802.16 (WiMAX), and Dedicated Short Range Communications (DSRC). The maturity of these technologies are one of the keys for enabling vehicle platooning. The wireless communication between vehicles that get involved in platooning forms a vehicular \textit{ad hoc} network (VANET), which is our primary focus in this project.

In VANETs, the IEEE 802.11p standard is emerging as a popular standard for such applications. The IEEE 802.11p standard is an amendment to the IEEE 802.11 standard which was proposed for Wireless Access in Vehicular Environments (WAVE). 
The IEEE 802.11p is licensed from 5.850 GHz to 5.925 GHz (the so called ``5.9 GHz band'') in the United States \cite{gao2016empirical}.
The standard is basically designed to enable communication in mobile environments e.g. V2V and V2I communication. The IEEE 802.11p physical layer is similar to that of the IEEE 802.11a standard. However, to cope with the high-speed vehicles, the bandwidth of the IEEE 802.11p is reduced to the half of the IEEE 802.11. Its Medium Access Control (MAC) protocol is inspired by the MAC of the IEEE 802.11e standard with some modifications to make it more suitable for mobility. Similar to the IEEE 802.11e, the IEEE 802.11p uses the Enhanced Distributed Channel Access (EDCA) mechanism. The EDCA mechanism allows the high-priority traffic to have a higher chance of acquiring the channel access and being transmitted compared to the low-priority traffic \cite{al2016wave}. 

Besides the IEEE 802.11p, WAVE also contains the standard of IEEE 1609, which is the upper layer standard.
IEEE 1609 completes the WAVE by its sub-detail standards, for instance, 
IEEE 1609.2 standard is responsible for the communication security; 
IEEE 1609.3 standard covers the WAVE connection setup and management. 
IEEE 1609.4 standard that is based on the IEEE 802.11p Physical (PHY) layer and Medium Access Control (MAC) layer supplies operation of high-level layers across multiple channels.
As shown in Table \ref{tab:layers}, within IEEE 802.11, DSRC is known as IEEE 802.11p, which amend the IEEE 802.11 on MAC and PHY layers \cite{libo10}. 

\begin{table}
    \centering
      \begin{tabular}{|l|l|l} 
        \cline{1-2}
        Application & Layer 7 
        & \rdelim\}{5}{3mm}[\texttt{IEEE P1609} (WAVE)] \\     \cline{1-2}
    \rowcolor[gray]{.9}    Presentation & Layer 6 \\    \cline{1-2}
        Session & Layer 5 \\    \cline{1-2}
    \rowcolor[gray]{.9}     Transport & Layer 4 \\  \cline{1-2}
        Network & Layer 3 \\   \cline{1-2}
    \rowcolor[gray]{.9}     Data Link & Layer 2 
    & \rdelim\}{2}{3mm}[\texttt{IEEE 802.11} (DSRC)] \\  \cline{1-2}
        Physical & Layer 1 \\   \cline{1-2}
      \end{tabular} 
      \caption{IEEE P1609 and IEEE 802.11 in OSI model}
    \label{tab:layers}
\end{table}

\section{Routing Protocols in VANET} \label{chpt:routing}
\subsection{Classification of VANET Routing Protocols}
\begin{figure}[h]
    \centering
    \begin{tikzpicture}[sibling distance=8em,
    	every node/.style={shape=rectangle, rounded corners, draw, align=center,
    				top color=white, bottom color=blue!20}
    	]
    	\node (n0) {Routing Protocols}
            child {node {Position \\ Based}}
            child {node {Broadcast \\ Based}}
            child {node {Cluster \\ Based}}
            child {node {Geo-Cast \\ Based}}
            child {node {Topology \\ Based}}
        ;
    \end{tikzpicture}
    \caption{Classification of Routing Protocols}
    \label{fig:routing_class}
\end{figure}
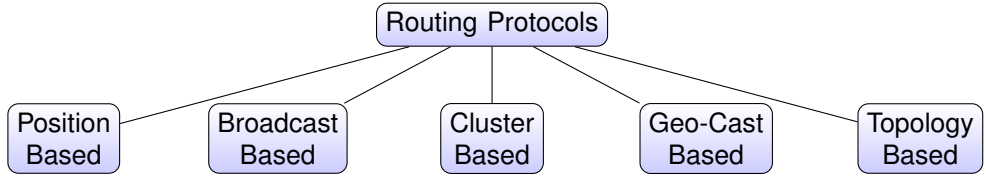

Routing is the process of selecting a path for traffic in a network, or between or across multiple
networks. Routing is usually performed by a dedicated device called a router. 
As we mentioned in Chapter 1, due to the high mobile and dynamic nature of VANETs, the reliability and efficiency of communication between vehicles critically relies on routing protocol, and various types of routing protocols have been proposed by researchers. They can be characterized into different classifications, as in \cite{ghori2018vanet}, \cite{paul2012vanet}, \cite{siddiquicritical}, \cite{dey2016review}, etc. 
Here we adopt a widely acknowledged one, as show in Figure \ref{fig:routing_class}, in which the VANET routing protocols are divided into five categories.

\subsubsection{Position Based Protocols}
Position based protocols requires the Global Positioning System (GPS) assistance to obtain the location of the vehicle as well as that of its neighbors. A source communicates with the destination by using geographical positions as well as its network address, and does not need any routing table. Therefore, it is suitable for highly dynamic environment. However, the satellite signal could become weak when the vehicle gets into the area such as tunnels. Examples of such protocols include the \textit{Greedy Perimeter Coordinator Routing} (GPCR) \cite{lochert2005gpcr}, \textit{Greedy Perimeter Stateless Routing} (GPSR) \cite{karp2000gpsr}, \textit{Geographical Source Routing} (GSR) \cite{lochert2003gsr}, etc.

\subsubsection{Broadcast Based Protocols}
Broadcast is a frequently used routing method in VANETs,
such as sharing traffic, weather, emergency, road condition among vehicles, and delivering advertisements and announcements. Broadcast is also used in unicast routing protocols (routing discovery phase) to find an efficient route to the destination. When the message needs to be disseminated to the vehicles beyond the transmission range, multi-hop is used.

The simplest way to implement a broadcast service is flooding in which each node re-broadcasts messages to all of its neighbors except the one it got this message from. Flooding guarantees the message will eventually reach all nodes in the network. 

Broadcast based protocols use broadcasting mechanism, i.e., flood data packets to all the available nodes in the entire VANET. Different relay selection techniques are used to reduce the message overhead. Examples of broadcast base protocols include the \textit{BROADCOMM Routing} \cite{lee2010survey}, \textit{Distributed Vehicular Broadcast Routing} (DV-CAST) \cite{tonguz2010dv}, \textit{Position Aware Reliable Broadcasting Protocol} (POCA) \cite{fasolo2005smart}, \textit{Density Aware Reliable Broadcasting Protocol} (DECA) \cite{nakorn2010deca}, \textit{Urban Multi-Hop Broadcast protocol} (UMB) \cite{korkmaz2004urban}, etc.

\subsubsection{Cluster Based Protocols}
Cluster based protocols divide the network into clusters so that each cluster consists of vehicles with same characteristics such as the speed and direction. If a vehicle node needs to communicate with a node within the cluster then the data will follow a direct path as it is considered to be a local communication. However, if the vehicle node wants to communicate a node which is outside the cluster, it must use its cluster head for reaching the destination. \textit{Clustering for Open Inter Vehicular Communication Network} (COIN) \cite{blum2003mobility} is the prime example of this type of protocols.

\subsubsection{Geo-Cast Routing}
Geocast refers to the delivery of packets to a group of destinations in a network identified by their  geographical locations.
In geo-cast routing protocols, the primary objective of geo-cast routing is to deliver the packet from a source node to all other nodes with a specified geographical region, which is defined as the \textit{Zone of Relevance} (ZOR). 
Most geo-cast routing methods are implemented as a multicast service within the ZOR.
However, to override the simple flooding of the geo-cast message from the source to the ZOR, a forwarding area called \textit{Zone of Forwarding} (ZOF) is used to
restrict the flooding inside it.
Examples of geo-cast routing include
\textit{GeoGRID} \cite{liao2000geogrid},
\textit{Inter-Vehicle Geo-cast Routing} (IVG) \cite{bachir2003multicast}, \textit{RObust VEhicular Routing} (ROVER) \cite{kihl2007reliable}, \textit{Distributed Robust Geo-cast Routing} (DRG) \cite{joshi2007distributed}, etc.

\subsubsection{Topology Based Protocols}
\begin{figure}[h]
    \centering
    \begin{tikzpicture}[
         level 1/.style={sibling distance=5cm},
         level 2/.style={sibling distance=10em},
         level 2/.style={sibling distance=4em},
    	every node/.style={shape=rectangle, rounded corners, draw, align=center,
    				top color=white, bottom color=blue!20}
    	]
    	\node (n0) {\textit{Ad-hoc} Routing Protocols}
            child {node {Proactive} 
                child {node {DSDV}}
                child {node {OLSR}}
            }
            child {node {Reactive} 
                child {node {DSR}}
                child {node {AODV}}
                child {node {TORA}}
                child {node {DYMO}}
            }
            child {node {Hybrid} 
                child {node {ZRP}}
                child {node {FSR}}
            }
        ;
    \end{tikzpicture}
    \caption{Classification of \textit{Ad-hoc} Routing and Exemplary Protocols}
    \label{fig:adhoc_class}
\end{figure}
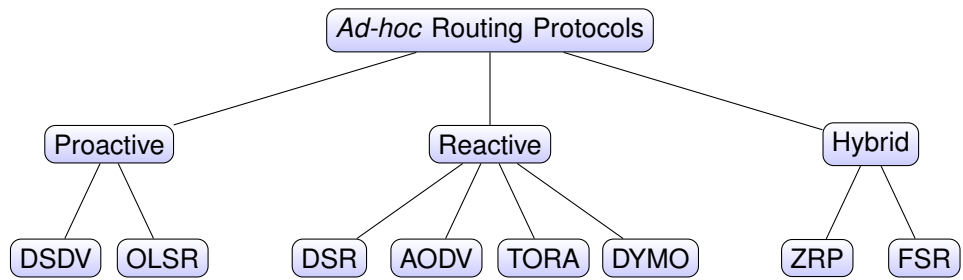

Topology based routing protocols use links information which stored in the routing table that exists in the VANET to forward data packet from source to destination. 
The topology based routing is usually further categorized into three types, the proactive or table-driven protocols, the reactive or on-demand protocols, and hybrid protocols. 
Figure \ref{fig:adhoc_class} illustrates this classification and provides several common routing protocol examples for each category.

\paragraph*{Proactive Routing Protocols}
In proactive routing protocols, each node continuously calculates and update the routes to every other node in the network. Routing information is periodically transmitted throughout the network for each node to maintain the routing table consistency, thus the proactive scheme is also called table-driven routing. When needed, a proactive protocol immediately provides the required routes to all reachable nodes in the network, with the minimal initial delay for transmission.
However, for highly dynamic network topology, the proactive schemes require a significant amount of signaling traffic and power consumption to keep routing information up-to-date and reliable. 
Examples of certain proactive routing protocols are \textit{Destination-Sequenced Distance Vector} (DSDV) \cite{perkins1994dsdv}, Optimized Link State Routing (OLSR) \cite{clausen2003olsr}, etc.

\paragraph*{Reactive Routing Protocols}
In reactive routing scheme, the protocol attempts to discover route only ``on-demand''.
For this purpose, a node initiates a route discovery process through the network. This process is completed once a route is determined or all possible permutations have been examined. Once a route has been established, it is maintained by a route maintenance phase until either the destination becomes inaccessible along every path from the source or until the route is no longer desired. In Reactive schemes, nodes maintain the routes to active destinations. A route search is needed for every unknown destination. 
This type of protocols reduce signaling traffic overhead thus save bandwidth and power consumption, but at the cost of increased latency when no route to the destination is available.
Common examples of reactive routing protocols are \textit{Dynamic Source Routing} (DSR) \cite{johnson1996dynamic}, \textit{Ad-Hoc On-Demand Distance Vector} (AODV) \cite{perkins2003aodv}, \textit{Temperarily Ordered Routing Algorithm} (TORA) \cite{park1997tora}, \textit{Dynamic Manet On Demand} (DYMO) \cite{sommer2007dymo}, etc.  

\paragraph*{Hybrid Protocols}
Hybrid protocols are composites of reactive and proactive routing, in the hope of taking advantage of both schemes. The main idea behind such protocols is to initiate route-discovery on demand but at a limited search cost. Common protocols based on hybrid approach includes \textit{Zone Routing Protocol} (ZRP) \cite{haas2002zone}, \textit{Fisheye State Routing} (FSR) \cite{pei2000fisheye}, etc.

\subsection{Review of Common Routing Protocols in VANETs}

\subsubsection{DSR}
\textit{Dynamic Source Routing} (DSR) is a reactive routing protocol based on source routing, which means that the source node explicitly lists the complete path to the destination in the header of each packet it sends. Therefore, each host in the network needs to determine the best route to others.
DSR consists of two phases: \textit{route discovery} and \textit{route maintenance}. In \textit{route discovery} phase, the source node broadcasts to all its neighbors within a wireless transmission range a Route Request (\texttt{RReq}) packet, which contains: request ID (in order to eliminate duplicate), target destination's address, and the route record. 
Upon receiving \texttt{RReq}, the destination node generates a Route Reply (\texttt{RRep}) packet that includes a list of addresses received in \texttt{RReq} and transmits it back through the same path to the source. \textit{Route maintenance} is the mechanism for detecting any broken route, which is required in DSR due to lack of periodic routing updates. When a node determines a problem at a hop, it will send Route Error packet which contains the addresses of both ends of the problem hop to the source. The source then can try to use any other known route to the destination in cache, or it can discover again to find a new path. All nodes get involved would need to update their route cache and send broadcast packets for discovering new routes, thus DSR could cause the so-called ``broadcast storm'' problem when the network topology changes.

There are several ways of optimization can be done in DSR to reduce the amount of overhead packets. For example, \texttt{RRep} can be piggybacked on another \texttt{RReq} packet sent to the targeted source. Another optimization is to allow intermediate node that receives \texttt{RReq} to respond with \texttt{RRep} if it knows a route to destination, furthermore, each node will delay for a time period proportional to its distance to destination before replying from its cache, to avoid the congestion caused by many intermediate nodes replies. Also, exponential backoff can be used in network partitions, so that the rate of initiating route discoveries can be limited and the ``broadcast storm'' can be reduced.

\subsubsection{DSDV}
\textit{Destination-Sequenced Distance Vector} (DSDV) is an extension of distance vector routing, which relies on the distributed Bellman-Ford algorithm to find the minimum-cost spanning tree. DSDV is a proactive table-driven routing scheme for VANET, its basic idea is that the destination nodes rather than intermediate ones provide the route freshness indication. In order to solve the famous distance vector problem, i.e., ``count-to-infinity'',  sequence number is added in each route entry. 
Each node preserves a routing table which stores a route to every possible destination in the network as well as the distance (i.e., the number of hops) to the destination. Each such entry in the table is marked with a sequence number assigned by the destination node.

When a node moves to a new location, it discovers its new neighbors, and then sends new distance vector with higher sequence number which indicates the freshness. 
The route with the most recent sequence number is always used, and among those with the same sequence numbers, the route with fewest number of hops is used. 
Eventually, the new routing information can be propagated to other nodes and flooded to the entire VANET. 
In DSDV, any updates in the routing tables are periodically broadcast in the network to maintain table consistency. The amount of traffic generated by these updates can be huge, especially at high mobility.

\subsubsection{TORA}
\textit{Temperarily Ordered Routing Algorithm} (TORA) is a loop-free and highly adaptive distributed routing algorithm based on the concept of link reversal routing. 
In TORA, when a node needs a route to a destination, it broadcasts a \texttt{query} packet which contains the address of the destination. The \texttt{query} packet is propagated until it reaches the destination or an intermediate node with a route to the destination. The corresponding destination or intermediate node then broadcasts an \texttt{update} packet listing its height to destination. As the \texttt{update} packet is propagated through the network, each node receives it needs to update its height. In this way, each node constructs the direct acyclic graph (DAG).
In TORA, when an invalid route is discovered, the node will look at all its neighbors to find a neighbor with the smallest height, and transmit an \texttt{update} packet. 
And if a node detects a network partition, it will generate a \texttt{clear} packet, reset the routing state, and removes invalid routes. 
The advantage of TORA is that it allows a route to all presented node in the network and the disadvantage is that the route maintenance is difficult in VANET. Whenever there is a change in the network, TORA broadcasts the \texttt{update} packets, which will result in high routing overhead.

\subsubsection{AODV}
\textit{Ad-Hoc On-Demand Distance Vector} (AODV) uses a modified version of broadcast route discovery mechanism from DSR. Instead of source routing, however, AODV relies on dynamically establishing route table entries at intermediate nodes. Like in DSDV, sequence number is included when updating routes. But each node maintains a monotonically increasing sequence number counter to supersede stale cached routes, which is different from DSDV. Thus AODV is often described as a combination of DSR and DSDV. And it is claimed to be the most commonly used routing protocol in VANET \cite{ghori2018vanet}.

In AODV, when a source node tries to send data to the destination node, it starts the route discovery process by broadcasting Route Request (\texttt{RReq}) to all its neighbors. Upon receiving \texttt{RReq}, if the neighbor nodes lack information pertaining to the request, they forward the packets to their neighbors. 
This process continues until \texttt{RReq} reaches either the destination node or the node who knows the path to destination, which will then send back to the source node a Route Reply \texttt{RRep} packet containing the number of hops to destination as well as its most recently seen sequence number from destination.
Every node that forwards \texttt{RReq} will cache a reversed route for itself back to the source, and every node that forwards \texttt{RRep} will create a forward route to destination. In this way, each node only needs to specify the next hop, rather than the entire route as in DSR. Note that the \texttt{RRep} packets are sent through the reversed path. And when a link is down, the upstream nodes will receive unsolicited \texttt{RRep} containing an infinite metric for the destination, the source node then needs to rediscover a new route.

\subsubsection{OLSR}
Optimized Link State Routing (OLSR) is a protocol that is based on traditional link state algorithm in which every node maintains the network topology information by exchanging link state information with others periodically. The advantage of this protocol is that it applies the MPR (Multipoint Relays) strategy to reduce the size of control information as well as the amount of rebroadcasting nodes. 

In OLSR, each node periodically broadcasts its \texttt{HELLO} messages containing information about its neighbors with bidirectional link as well as other known neighbors. The exchange of these \texttt{HELLO} messages between (1-hop) neighbors permits each node to learn the knowledge of its neighbors up to 2 hops. Each node then maintains a neighbor table, which records the information about its 1-hop neighbors as well as the link status (which could be unidirectional, bidirectional, or MPR), and a list of 2-hop neighbors that these 1-hop neighbors give access to. Each entry in the neighbor table has an associated holding time, upon expiration of which that entry is no longer valid and thus removed. The neighbor table also has a sequence number, which is incremented every time a node selects or updates its MPRs. We will illustrate the procedure of MPR selection in Section \ref{sec:mpr}.

TC messages are sent periodically by each node ($N$) to declare its MPR Selector set, i.e., a list of neighbors who have selected the sender node $N$ as a MPR. A sequence number is also associated with the TC message. A node with an empty MPR Selector set (i.e., none of its 1-hop neighbors has selected it as an MPR) may not generate TC messages. In OLSR, only MPRs are allowed to generate TC messages, and only MPRs are responsible for retransmission of TC messages.

The information diffused in the network by TC messages helps each node to build its topology table. Each node maintains a topology table, and upon receiving a TC message, the following procedure may be executed:
\begin{itemize}
    \item if there exists an entry of the topology table has the same destination with TC:
        \begin{itemize}
            \item if the topology table has higher sequence number, the TC message will be ignored;
            \item if the TC message has higher sequence number, then the corresponding entry in the topology table will be replaced by a new entry recorded from the TC message;
            \item if the topology table and the TC message have the same sequence number, then the holding time of that entry will be refreshed;
        \end{itemize}
    \item otherwise (i.e., all entries in topology table have different destination from TC), a new entry is recorded in the topology table.
\end{itemize}

With the neighbor table and the topology table, each node in the network would calculate and maintain a routing table, which determines the route to other destinations. The entries in the routing table consist of the destination address, next hop address, as well as the distance (i.e., number of hops) to destination. If any change occurs in either the neighbor table or the topology table, the routing table is re-calculated to update the route information about each known destination in the network.

\subsection{Multipoint Relays (MPRs)} \label{sec:mpr}

\begin{figure}
    \centering
    \begin{subfigure}[b]{0.45\textwidth}
        \includegraphics[width=\textwidth]{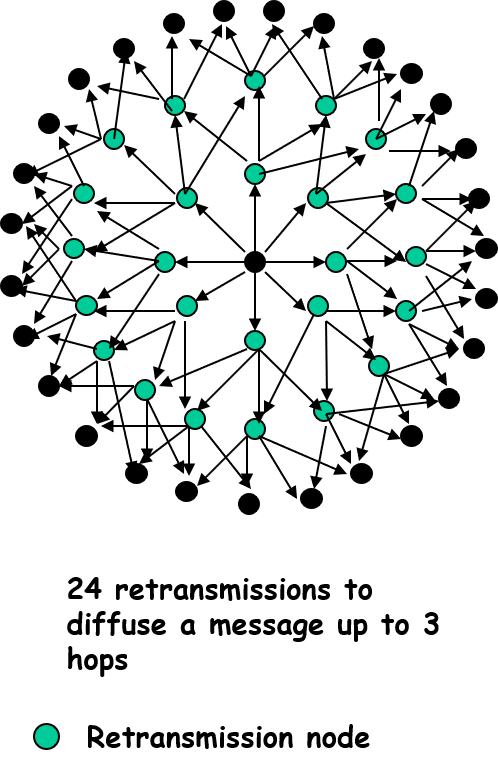}
        \caption{Regular Flooding}
        \label{fig:flood}
    \end{subfigure}
    ~ \quad
    \begin{subfigure}[b]{0.45\textwidth}
        \includegraphics[width=\textwidth]{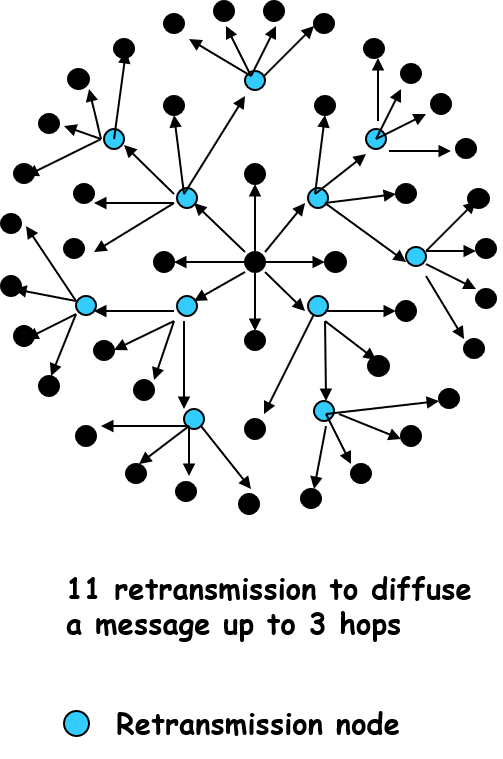}
        \caption{Flooding using MPRs}
        \label{fig:mpr}
    \end{subfigure}
    \caption{Comparison of flooding a packet in a network} 
    \label{fig:olsr}
\end{figure}

The idea of multipoint relays (MPRs) is to reduce the flooding of packet by reducing the duplicate retransmission of broadcast packets. 
Unlike in regular flooding where each node is responsible to retransmit a packet it receives, only selected MPRs are allowed for retransmission. Therefore, the route is a sequence of hops through the MPRs from source to destination. As in Figure \ref{fig:mpr}, the blue circles represents MPRs, and only them are allowed to retransmit the packet originated from the center node. 

Each node ($N$) in the network selects and maintains its own MPRs from its 1-hop neighborhood, the MPRs are selected such that it covers all symmetric 2-hop neighbors of $N$. Those 1-hop neighbors of $N$ which are not selected as its MPR receives and processes messages from $N$, but are not allowed to retransmit broadcast messages gathered from $N$.

The MPRs of $N$, denoted as $MPR (N)$, is then an arbitrary division of the symmetric 1-hop neighborhood of $N$ which comply with the following aspect: every node in the symmetric strict 2-hop neighbors of $N$ must have a symmetric link towards $MPR (N)$. The tinier the MPRs, the less control traffic overhead outcome from the routing protocol. Respective node also takes care of information about the neighbors that have selected it as an MPR. This set is called the MPR Selector set. A node obtains the information from recurring \texttt{HELLO} messages received from its 1-hop neighbors. 
The MPRs change over time as the topology of network changes (i.e., when a node selects another MPR set) and will soon be indicated by the selector nodes in their \texttt{HELLO} messages. 

Figure \ref{fig:olsr} shows the retransmission of diffuse a message up to 3 hops. We can see that using MPRs could reduce 24 retransmissions down to 11. 

\subsection{Literature Review on Analysis of VANET Routing Protocols}

There are many comparative analysis of VANET routing protocols in the literature, the compared protocols and the simulation environment vary a lot, and the results and conclusions are quite different.
Here we only list a few of recent studies, mainly on some of the topology based VANET routing protocols we mentioned above.

\cite{hamid2015performance} analyzed the performance of OLSR, AODV and DSDV protocols in a VANET network and examined the impact of varying mobility, density and pause time on the functionality of these protocols. They demonstrated that AODV have better performance in terms of throughput and packets delivery rate, whereas OLSR have best performance in terms of packet delivery . 

In \cite{ahmed2014performance}, the authors evaluated AODV, DSR and OLSR performance in urban scenarios under varying metrics such as node mobility, vehicle density, and with varying traffic rates. The results showed that clustering effects created by cars aggregating at intersections have remarkable impacts on evaluation and performance metrics, and OLSR performs best in most of the simulated traffic situations. 
\cite{chouhan2015analysis} compared DSDV, OLSR and AODV using NS-3, and established that among the routing protocol used, OLSR proves to be the best in terms of highest packet delivery rate and lowest packet loss rate for every node, also has the least overhead.

\cite{sallum2018performance} also compared DSDV, OLSR and AODV using NS-3, but under several different environment settings, they showed that the DSDV and OLSR have better performance than AODV for low-density and low-speed scenarios, and as the network density or nodes' velocity increases, the OLSR protocol performs better than the other two. 

\cite{santoso2012performance} came up with different conclusion, in which the performance of AODV, DSDV and OLSR were compared and evaluated in a safety application scenario using NS-3, and the simulation results show that DSDV could adapt to the requirement of safety application consistently, while OLSR and AODV failed to achieve the expected performance.

AODV has been reported to perform better than DSDV and OLSR in VANETs more often. For example, again using NS-3 and SUMO (Simulation of Urban Mobility), \cite{gupta2014performance} claimed that AODV gives best results in high mobility environment, after comparing with OLSR and DSDV. \cite{zhang2016assessing} drew the same conclusion after comparing with DSR. 
\cite{sallam2015performance} came out with the conclusion that AODV is far better than OLSR as it gives very good result with high speed vehicles.
Similar conclusion was drawn in \cite{phouthone2015simulation} by using NS-2. 
Another performance evaluation of topology based routing protocol (DSDV, AODV and LSGR) was done in \cite{kumar2016performance} and reported AODV as the best.

Other routing protocols have also been claimed to perform better.
In \cite{houssaini2017comparative}, the authors explored 8 different routing protocols, including proactive (DSDV, OLSR, FSR), reactive (AODV, DYMO, DSR), hybrid (ZRP) and positionbased protocols (GPSR), by using NS-2 with vehicle mobility under urban traffic generated from VanetMobiSim,
and concluded that the geographic routing protocols perform better in VANET since the position information can be useful. And \cite{setiabudi2016performance} concluded that hybrid protocols always work well in ad-hoc environment and specially with high speed nodes, after the comparative study between GPSR and ZRP.





\section{Design}
In this project, sensor and control packets will be transmitted over an emulation of the wireless vehicular ad-hoc network (VANET) that uses an efficient ad-hoc routing protocols. The emulation of the mobile VANET is implemented over a fixed network. The distributed \texttt{tux} machines at Auburn University are used to simulate the experiments. Wireless nodes are executed on different computers, and communicate with each other through the network.


\subsection{Vehicle Platooning Application}
A simple road train application program will be implemented to simulate the control for a lead truck and multiple following vehicles, which will enable them to participate in automatic platoon by providing the following functions.
\begin{itemize}
    \item Form a platoon behind a lead truck. 
A vehicle may initiate forming a platoon with a lead truck where the lead truck will determine the road train parameters, e.g. road train speed and direction. Once the platoon is formed, the road train control application will ensure that the following vehicle and the lead truck will travel at the correct speed to
maintain the platoon configuration, i.e. the distance of the gap between the following vehicle and lead truck is within a certain range (e.g., between 10 to 20 meters). 
The actual speed of each vehicle in a road train may be different. 
    \item Join an existing platoon.
A vehicle that receives vehicle sensor information from other vehicles in the road train may join the road train. Once it joins the road train, it will maintain the road train configuration as above, i.e. it must maintain a spacing from the front and back vehicle of a certain distance (e.g., between 10 to 20 meters).
    \item Leave a platoon. 
A vehicle may leave a road train and need not be bound by the rules for maintaining the road train configuration.
\end{itemize}

\subsubsection{Packet Propagation Loss Model}

\subsection{Vehicle Behavior and Packet Categories}
\subsubsection{Vehicle Information Packet Broadcasting}
Every 10 milliseconds, each vehicle will broadcast a packet which containing its own driving information. 
The datagrams of the vehicle information packets contain the following:
\begin{itemize}
    \item GPS position
    \item Velocity
    \item Acceleration
    \item Brake control
    \item Gas throttle
\end{itemize}

Upon receiving the vehicle information packets from other vehicles, each vehicle would check if the incoming packet is the newest one it has received from the source vehicle (by checking the sequence number). If that is the case, it would update the its own knowledge of the source vehicle if necessary. Also, by checking its MPR selector table, it will know whether to forward/retransmit the received packets with the newest sequence number to all its neighbors, aka every other vehicle in its transmission range (i.e., 100 m), except the one from which it receives the packet (i.e., the previous hop) or the source. The retranmission will be performed if the receiver is selected as an MPR of the previous hop.

\subsubsection{Platooning Behavior and Related Packets}
The CACC system needs to be simple, reliable and trustworthy, its primary objective is to maintain a
certain gap in the road train. 
Each following vehicle could join or leave the road train leading by the lead truck by sending \texttt{JOIN} or \texttt{LEAVE} requests. The design of the joining and leaving procedures is one of the key building blocks in our project.

The lead truck sends its driving information to the following vehicles so that the following vehicles in the road train can duplicate its behavior and stay in the road train. The following vehicles send their driving information to the lead truck, so that the lead truck can keep track of their positions, and allow other vehicles to join or leave the road train. The sequence number and the IP address are included in the packet header, and in addition to those \texttt{NORMAL} packets for exchanging the vehicle information, we further define five different types of special packets for controlling the platooning behavior, they are \texttt{JOIN}, \texttt{LEAVE}, \texttt{ACK\_JOIN}, \texttt{NOTIFY}, and \texttt{OK}. 

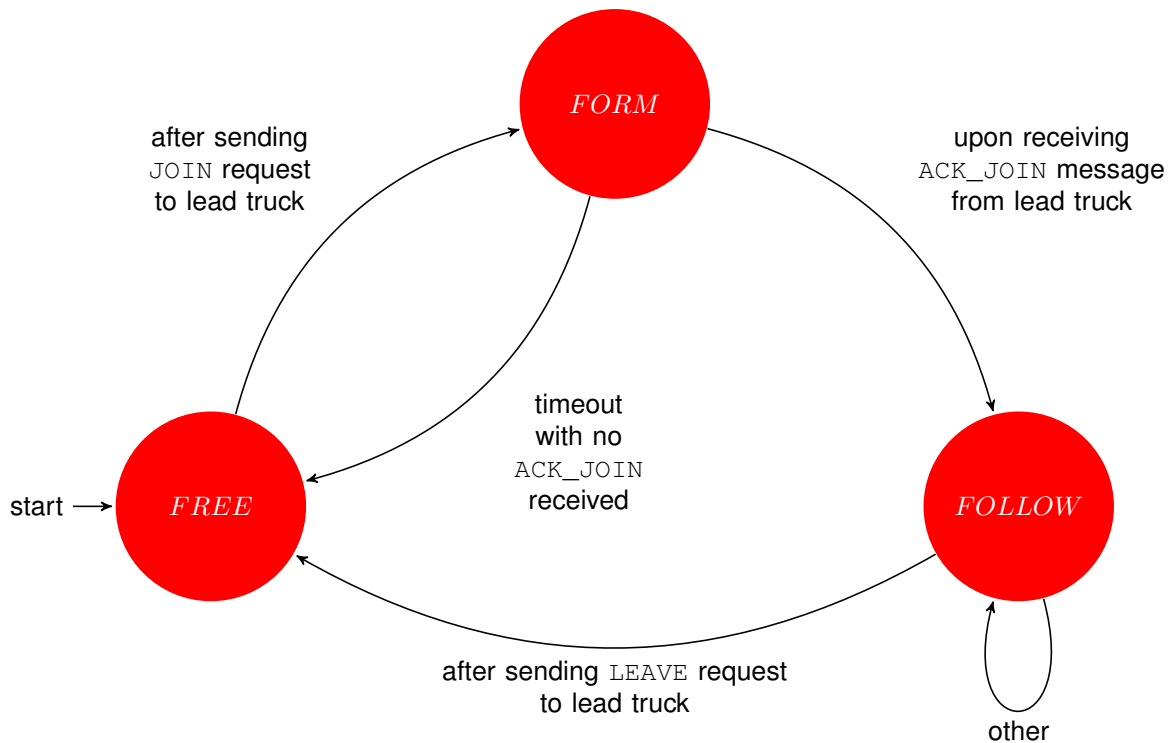
\begin{figure}
    \centering
    \begin{tikzpicture}[->,>=stealth',shorten >=1pt,auto,node distance=7.5cm, semithick]
      \tikzstyle{every state}=[fill=red,draw=none,text=white,minimum size=2.5cm]
    
      \node[initial,state] (A)                    {$FREE$};
      \node[state]         (B) [above right of=A] {$FORM$};
      \node[state]         (C) [below right of=B] {$FOLLOW$};
    
      \path (A) edge [bend left] node { \begin{tabular}{c} 
                            after sending \\ \texttt{JOIN} request \\ to lead truck  \end{tabular} } (B)
            (B) edge [bend left] node { 
                            \begin{tabular}{c} 
                            timeout \\with no \\ \texttt{ACK\_JOIN} \\received  \end{tabular} } (A)
                edge [bend left] node { \begin{tabular}{c} upon receiving \\ 
                            \texttt{ACK\_JOIN} message \\ from lead truck  \end{tabular} } (C)
            (C) edge [bend left] node { 
                     \begin{tabular}{c} after sending \texttt{LEAVE} request\\ to lead truck \end{tabular} } (A)
                edge [loop below] node {other} (C);
    \end{tikzpicture}
    \caption{Platooning state transitions of vehicles (other than lead truck)}
    \label{fig:sm}
\end{figure}

The following vehicles can send \texttt{JOIN} or \texttt{LEAVE} request to the lead truck, and the lead truck should reply an \texttt{ACK} for that request. 
The \texttt{JOIN} request can only be sent when a following vehicle is not in the road train, i.e., its vehicle state is \textit{FREE}. After it sends a \texttt{JOIN}, the vehicle state becomes \textit{FORM}, which indicates that this vehicle is waiting for the \texttt{ACK\_JOIN} message from the lead vehicle. 
To avoid collision while merging into the middle of the platoon, it is only allowed to join the road train when the lead truck makes sure other vehicles have prepared enough space for it to merge and sends it an \texttt{ACK\_JOIN} message. After it receives \texttt{ACK\_JOIN} and replies with \texttt{ACK\_JOIN}, the vehicle state becomes \textit{FOLLOW}, which indicates that this vehicle has joined the road train, and its driving is controlled by following another vehicle in the road train. It is only allowed to make \texttt{LEAVE} request when it is in the road train, i.e. the vehicle state is \textit{FOLLOW}.
This is like a state machine (See Figure \ref{fig:sm}). Note that in our design, the \texttt{ACK\_JOIN} 
messages sent by the lead truck and by the following vehicle have totally different functionality, the details are described in Section \ref{sec:join}.
While the connection established, the lead and following vehicles communicate their driving information via \texttt{NORMAL} packets.

\subsubsection{Join the Platoon} \label{sec:join}
The specific procedure for the following vehicles to join the road train is as follows:
\begin{itemize}
    \item Initially, only the lead truck stays in the road train. The following vehicles could send request to join the road train at any time if it is not yet in. 
    \item The lead truck then checks the information of the following vehicle that contains in the packet it receives, and determines whether to accept or decline the request of joining the road train, based on if the following vehicle is at the same direction. Once the lead truck decides the following vehicle could have the chance to join the road train, it would then determine which position in the road train the requesting vehicle should join, that is, put the requesting vehicle into the road train right after the nearest vehicle ahead of it, based on the vehicles' current GPS information. If no vehicle in the road train is ahead of it, then it should slow down and follow the lead truck.
    \item If the requesting vehicle were asked to join to the end of the road train, it will then received an \texttt{ACK\_JOIN} message from the lead truck, which tells it which specific vehicle it should follow. If the lead truck determines that the requesting vehicle to join in the middle of the road train, it will inform other vehicles to leave enough space for the joining vehicle first. Only after the vehicle it informed replied an \texttt{ACK} message when enough space has been maintained and thus it is safe to join, the lead truck then sends an \texttt{ACK\_JOIN} to the requesting vehicle.
    \item Once the following vehicle which initiated a \texttt{JOIN} request receives an \texttt{ACK\_JOIN} from the lead truck, it would then accelerate to catch up the platoon, or slows down for waiting the platoon, until reaches a desired location so that it keeps an appropriate distance from the vehicle in front of it in the road train, it would then moves to the right lane of the road, set the speed and acceleration in order to follow the vehicle in front of it, upon the information received from that vehicle most recently. Remark that to simplify the model, we assume the acceleration/de-acceleration to the desired speed for catching up or slow down is instantaneous, so is the braking and moving to the right lane and joining the platoon.
    \item After joining the road train, the requesting vehicle sends an \texttt{ACK\_JOIN} to the lead truck. The lead truck then update its own knowledge about the road train topology, and could be able to process the next \texttt{JOIN} or \texttt{LEAVE} request.
\end{itemize}

\begin{figure}[htb!]
    \centering
    \includegraphics[width=\textwidth]{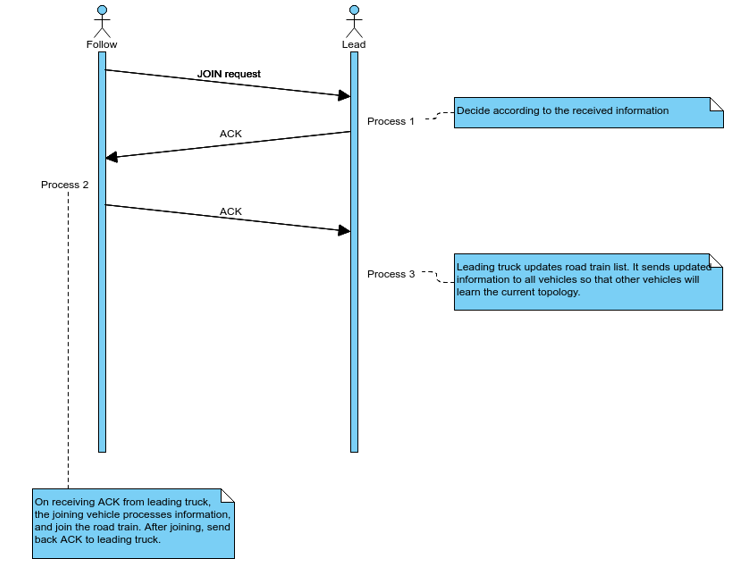}
    \caption{Procedure for a following vehicle joining the road train to the end}
    \label{fig:j1}
\end{figure}

\begin{figure}[htb!]
    \centering
    \includegraphics[width=\textwidth]{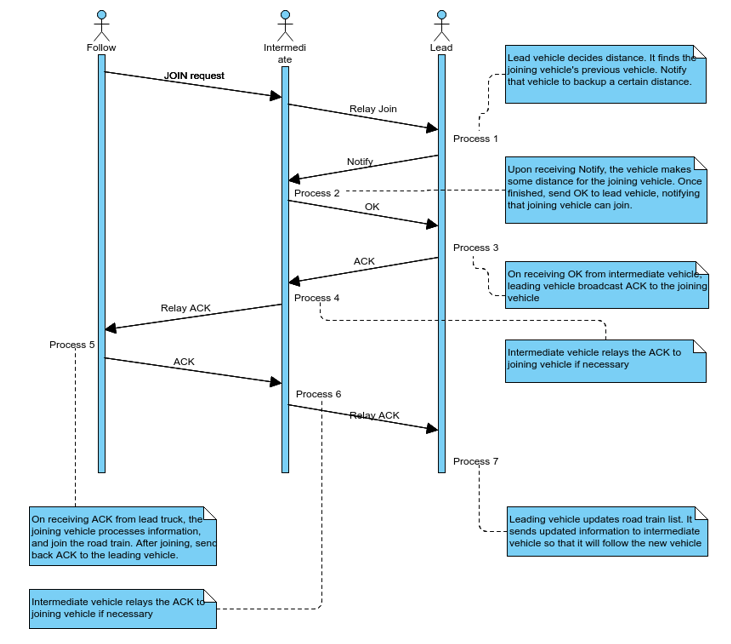}
    \caption{Procedure for a following vehicle joining the middle of the road train}
    \label{fig:j2}
\end{figure}

The above procedures are depicted in Figure \ref{fig:j1} and Figure \ref{fig:j2}, which illustrate different situations whether the following vehicle is about to join to the end or middle of the road train, respectively.

\subsubsection{Leave the Platoon}
For a following vehicle to leave the road train, the procedure is much simpler. The following vehicle $k$ sends a \texttt{LEAVE} message to the lead truck, without the need of receiving any response, it then moves to the left lane, and is no longer be controlled by the road train. To increase the reliability of the communication, we let the leaving vehicle sends \texttt{LEAVE} message for three times. Once the lead truck receives the \texttt{LEAVE} packet, it needs to inform the vehicle ($k+1$) in the road train which was following the left vehicle $k$ to follow the one previously was ahead the left one, (i.e., $k+1$ should follow $k-1$ instead of $k$ after $k$ leaves), by sending a \texttt{NOTIFY} message to vehicle $k+1$. This information can be calculated from the road train topology table cached in the lead truck, and upon receiving the \texttt{LEAVE} message, the lead truck need to update its knowledge about the road train topology.
The informed vehicle, as well as all other vehicles behind it, then speed up a little and fill up the space. 
Note that the only information that the lead truck notifies the informed vehicle is which vehicle in the road train it should follow, and the informed vehicle changes its speed based on its own knowledge about other vehicles' driving information from corresponding \texttt{NORMAL} packets it received. 

\subsection{Packet Forwarding Protocols}
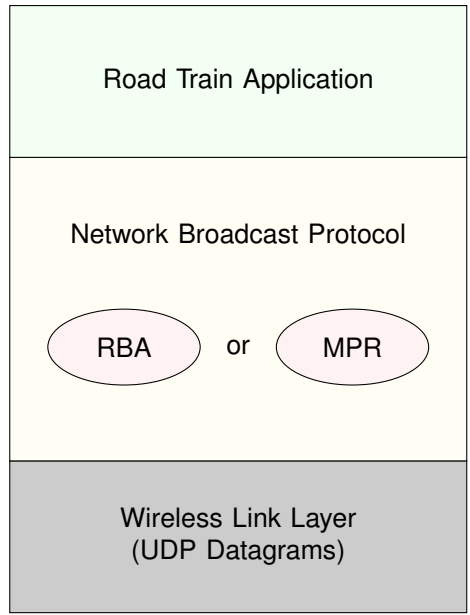
\begin{figure}
    \centering
    \begin{tikzpicture}
        \filldraw[fill=gray!40!white, draw=black] (0,0) rectangle (6, 2) node[pos=.5] { 
            \begin{tabular}{c} Wireless Link Layer \\ (UDP Datagrams)
            \end{tabular}
        };
        \filldraw[fill=yellow!5!white, draw=black]  (0,2) rectangle (6, 6);
        \node at (3, 5) {Network Broadcast Protocol};
        \filldraw[fill=red!5!white, draw=black] (1.5,3.5) ellipse (1 and .5) node{RBA};
        \filldraw[fill=red!5!white, draw=black] (4.5,3.5) ellipse (1 and .5) node{MPR};
        \node at (3,3.5) {or};
        \filldraw[fill=green!5!white, draw=black]  (0,6) rectangle (6, 8) node[pos=.5]{Road Train Application};
    \end{tikzpicture}
    \caption{Overview of the VANET Protocols Components}
    \label{fig:struct}
\end{figure}

In order for a vehicle to learn about the position, speed and acceleration of road train and other cars, each vehicle and the lead truck must transmit its sensor data containing its GPS position, speed and acceleration at a constant period of 10 to 100 milliseconds. These sensor data are then forwarded multi-hop. In this section, we will design two schemes of broadcasting, one uses MPRs while the other one does not. Both algorithms aims at reliably sending information packets to other vehicles. The network broadcast protocols will be built on top of the simple UDP (User Datagram Protocol) socket interface that emulates the wireless link layer. The structure is shown as in Figure \ref{fig:struct}. Note that the wireless link layer is not part of this project, but represents protocols that could conceptually be added or substituted.

\subsection{Reliable Broadcast Algorithm}
We first design a \textit{reliable broadcast algorithm} (RBA) on top of the link layer. 
Initially, each vehicle will generate a vehicle sensor packet depending on its dynamics and forwards it to all its neighbor nodes through its RBA layer. Eventually the packet will be forwarded through multiple hops and reach all other vehicle destination addresses. The RBA layer of the intermediate nodes performs the task of forwarding packets reliably from one node to another. The packet header must contain at least the following fields:
\begin{enumerate}
    \item Sequence number
    \item Source address
    \item Previous hop (address)
\end{enumerate}

Each vehicle node will run the program for the RBA as follows.  
In the RBA layer, each vehicle maintains a \textit{cache table} of the broadcast packets that it has received from every other source node in the network, with the largest sequence number of the packets that it has received from the source up to that time point, i.e. the \textit{cache table} contains information on broadcast packets that the node has forwarded previously. Now for a node with address $addr_i$, consider the case it receives a packet from another node with address $addr_{prev}$, which can be examined from the \textit{``Previous Hop''} field specified in the packet header,  it will first check the source address $addr_s$ of the received packet. If the packet source is itself ($addr_s=addr_i$), the packet will be discarded; otherwise it then compares the sequence number $sn_p$ of the packet with the largest sequence number $sn_s$ it cached for source $addr_s$ in its \textit{cache table}. 
If there is no $sn_s$ record in the \textit{cache table}, or the sequence number in the \textit{cache table} is smaller than that of the received packet, i.e. $sn_s<sn_p$, which means this packet is the newest from source $addr_s$ for node $addr_i$, and needs to be forwarded. The node with $addr_i$ will then update its cache table, in which the sequence number from source $addr_s$ is replaced by $sn_p$, and the \textit{broadcast number} $bn$ for that packet is set to 1. In addition, it changes the \textit{``PrevHop''} field of the packet to its own address $addr_i$, and forwards a copy of the packet to each of its neighbors except the one from which it received this packet, i.e., $addr_{prev}$. Eventually the packet is sent to all the reachable wireless nodes. 

To ensure the reliability of the broadcast protocol, we also design a \textit{re-broadcasting} mechanism for increasing the redundancy, in case the previous broadcast packet was lost. With \textit{re-broadcasting}, 
if the sequence number in the \textit{cache table} is the same as that of the received packet, i.e. $sn_s \ge sn_p$, the \textit{cache table} will also be updated, by adding one to the \textit{broadcast number} of that packet. And that packet may be forwarded depending on the probability determined by the \textit{broadcast number}, which in fact records the number of times the same packet has been re-broadcasted. If the packet has been broadcast once, there is a 50\% probability of re-broadcasting it to its neighbors; if it has been broadcasted twice, then the probability is 25\%, and so on.

\subsection{Broadcasting with MPRs}
Alternatively, a broadcast protocol using MPRs on top of the link layer will be designed and implemented. 
In this protocol, each vehicle continuing broadcasts \texttt{HELLO} messages which contain its own one hop neighbors as well as their link status every 20 milliseconds, and we assume that any other
vehicles could receive the packets only when their distance is within 100 m. The link status
of each one hop neighbor could be either uni-directional, bi-directional, or MPR.

Upon receiving \texttt{HELLO} messages from others, each vehicle is able to learn the knowledge of its neighbors up to two hops. After sensing the neighbors, the \texttt{HELLO} message receiver then could have enough information to select its MPRs from its bi-directional linked one hop neighbors, by knowing all its two hop neighbors (as well as their access through nodes). The \texttt{HELLO} message receiver then could update all the neighborhood information in its \textit{neighbor table}, which contains its one hop neighbors, the status of the link with these neighbors, and a list of two hop neighbors that these one hop neighbors give access to, as well as a sequence number which specifies the most recent MPR set that the local node keeping this neighbor table has selected. Every time a node selects or updates its MPR set, this sequence number is incremented to a higher value. Moreover, on reception of \texttt{HELLO} messages, each vehicle is able to know is this one hop neighbor has selected it as an MPR or not. Based on that, it can construct a MPR selector table, so that it will know whether to retransmit the received packets (other than \texttt{HELLO} messages). 

Every broadcast message coming from these MPR Selectors of a node is assumed to be retransmitted by that node. This set can change over time, which is indicated by the selector nodes in their \texttt{HELLO} messages.

\subsubsection{Neighbor Sensing and MPR Selection}
After receiving \texttt{HELLO} messages, the receiver vehicle (let us denote it as $R$) might need to update its neighborhood information. First of all, we need to check whether the sender (we denoted it as $S$) was $R$'s two-hop neighbors previously, it needs to be removed if it is the case. 
In any case, $S$ should then be added into $R$'s one-hop neighbors if it was not there, and the link will be set as uni-directional if $R$ has not been declared as $S$'s one hop neighbor in this \texttt{HELLO} message. On the other hand, in the \texttt{HELLO} message, if $R$ is already $S$'s one hop neighbor, then the link for $S$ will be set as bi-directional. 

Next, we need to update $R$'s two-hop neighbors information. Note that each of $R$'s two-hop neighbor ($M$) will maintain a set to record through which $R$'s one-hop neighbor, $M$ could get access through and reach $R$. Also note that the received \texttt{HELLO} contains the link status of all $S$'s one-hop neighbors, upon which $R$'s \textit{MPR selector table} could be updated. Moreover, if $S$ was already a bi-directional one-hop neighbor for $R$, then there might exist some $M$  which use $S$ as an access through point to get to $R$. But if now the bi-link between $S$ and $M$ was unlinked (as specified in the received \texttt{HELLO}), in that case, we need to remove $S$ from $N$'s access through set, and we need to traverse the two-hop neighbors and do this for all possible $M$'s. If any such $M$ has empty access through set after removing (i.e., $S$ was the sole access through point for $M$ to reach $R$), it will also be cleaned from $R$'s two-hop neighbors.

We also need to traverse for each of $S$'s one hop neighbor $N$, which could be a candidate for $R$'s two hop neighbor. 
If $N$ is already a two hop neighbor of $R$, we need to update $N$'s access through set by adding source if necessary. Otherwise, $N$'s access through set will be newly constructed and contains only $S$, with that information, $N$ is then added as a new two-hop neighbor for $R$.

If there is any change, $R$'s MPRs need to be selected again. We use heuristics for MPR selection. First we sort the two hop neighbors by the number of one-hop neighbors (called access-through set $\mathcal{A}$, in which the element is denoted as $A$) they connected with (note that the connection must be bidirectional). Then we allow the one with the fewest links with the access-throughs ($A$'s) chooses its MPR first, by randomly choose an element in $\mathcal{A}$; Then the one with second-fewest $A$'s; and so on, until all two-hop neighbors are traversed.

Lastly, we update the one-hop neighbors by labeling their link status. If a one-hop neighbor has been selected as an MPR, we label its link status as MPR. However, if a one-hop neighbor was an MPR previously, but has not been selected this time, we need to degrade its link status as bi-directional. 

\begin{table}[htb]
    \centering
    \begin{tabular}{|c|c|}
    \hline \hline
\multicolumn{2}{|c|}{Node 4's One-Hop Neighbors}  \\ \hline
Node ID & Status \\
\hline
   3   &   MPR  \\
    5   &    MPR   \\
    6   &    BI   \\
    7   &   BI  \\
    8   &   BI  \\
    9  &    BI  \\
\hline \hline
\multicolumn{2}{|c|}{Node 4's TWO-Hop Neighbors}  \\ 
\hline 
 Node ID &   Access Through (MPR) \\
\hline 
    1    &     3  \\
    2     &    3  \\
    10    &    5  \\ \hline \hline
    \end{tabular}
    \caption{Exemplary Neighbor Table}
    \label{tab:nbTab}
\end{table}

A typical neighbor table would be like in Table \ref{tab:nbTab}, which is obtain from our experiment for 10 vehicles in the VANET.

\subsubsection{OLSR Protocol for Special Packets}
Those 7 types of special packets for joining and leaving the road train are the only packets use the optimized linked state routing (OLSR) protocol, since they are the only packets that specify the destination. The OLSR protocol is based on MPR selection, which we have already illustrated. The MPRs will send topology control (\texttt{TC}) messages to all its one hop neighbors every 30 milliseconds, in which only its own MPR selector table is declared. The sequence numbers of \texttt{TC} just use the sequence number of the neighbor table in that vehicle, so even new \texttt{TC} messages are generated, if no neighborhood information is changed, the sequence number would not increase. 

The \texttt{TC} messages are used for every vehicles in the network to update their \textit{topology tables}. Each vehicle maintains its own \textit{topology table}, which has an entry for each of the other vehicles it knows in the network. Each entry contains some information about the initiator of the \texttt{TC} messages, which includes the address, MPR selectors, the largest sequence number of received \texttt{TC} messages from the initiator, and a holding time. Upon receiving an \texttt{TC} message, the corresponding entry is updated only when the \texttt{TC} message is larger than the one saved in the \textit{topology table}. And if the two sequence numbers are equal, the holding time of that entry will be refreshed.

The \textit{topology table} is used for updating the \textit{routing table}, from which the vehicle could calculate which the next hop is for sending or forwarding those 7 types of special packets. The calculation of next hop requires keeping track of a reversed path from the destination through its MPRs, to the MPRs' MPRs if necessary, until one of the vehicle's one hop neighbor. \textit{Bread First Search} is used for the calculation.

By doing above procedures, $R$'s neighborhood information as well as MPR selector table have been updated. If there is any change, the neighbor table sequence number will be increased, and the routing table will also be updated. The neighborhood information will be used for sending and forwarding packets, and also would be written into new \texttt{HELLO} messages. The MPR selector table would be transmitted to other vehicles if it is an MPR. 

\subsection{Network Configuration}

The status of each packet sending and receiving would be recorded into a configuration file. 
A configuration file helps us to look at the link state between the vehicles, also the status of each sent or received packet. The vehicles write information into the configuration file, but do not track and utilize information from it. The configuration file looks like as in Table \ref{tab:my_label}.

\begin{table}[h]
    \centering
    \begin{tabular}{c c c r r l}
    \hline  \hline
        Node 1 & tux055, & 10010 & 50 & 120 & links 2 \\
        Node 2 & tux055, & 10011 & 80 & 120 & links 1 3 4\\
        Node 3 & tux060, & 10010 & 180 & 170 & links 2 4\\
        Node 4 & tux060, & 10011 & 150 & 60 & links 2 3\\
         \multicolumn{6}{c}{$\cdots \cdots$} \\
    \hline
    \end{tabular}
    \caption{Exemplary Configuration File}
    \label{tab:my_label}
\end{table}

\subsection{Simulation set-up}
To simulate different number of vehicles in a road train, the program needs to run on different tux machines. As there is no physical access to tux machines, remote login is needed. Multiple remote login windows may be necessary. The lead truck should always be instantiated first, and when the program starts to run, the lead truck will first erase the configuration file if that exists, and then write its own information into the file. The configuration file helps to track the existing nodes in the simulation environment, as well as their link states. Only after the lead vehicle were added into the environment, other vehicles, aka following vehicles, could enter the simulation. Each of the vehicle would have a unique node ID, starting from 1 for the lead truck, and incrementally generated for every vehicles when they were added. Each vehicle continues broadcasting \texttt{HELLO} messages which can be received by all other vehicles within a certain transmission range (i.e., 100 meters) with a loss rate. Upon that, each vehicle could maintain a neighbor table and select MPRs. Each vehicle continues sending packets containing information about its status to all its one hop neighbors, among which only MPRs will retransmit the received vehicle information packets. In this way, all other vehicles are able to sense the information by receiving the packets if they are within the transmission range of any vehicle. Following vehicles can make \texttt{JOIN} and \texttt{LEAVE} request depending on if they are in the road train, through \textit{optimized linked state routing} (OLSR) protocol. Note that all packet are sent and received among vehicles by using use UDP protocol.

\section{Implementation}

\subsection{Vehicle Model}
A vehicle class is established to describe the attributes and behavior information of a vehicle, while the behavior of the lead truck and other following vehicles are quite different, we have two different subclasses to distinguish them. 
As they also have many functionality in common, \textit{Vehicle} will be implemented as an abstract class, from which two different classes \textit{LeadingTruck} and \textit{FollowingVehicle} are extended.
The \textit{utility} methods are written separately in a handler class for vehicle parameters, includes keeping track of their GPS, controlling of their appropriate driving behavior, performing I/O operations with the configuration file, etc.

\subsection{Socket Programming}
For networking communication, we directly use the UDP in \texttt{java.net}. To write and retrieve information in the packets, a network handler class is written. Since a vehicle object contains all information that we want to send, we exploit the \texttt{Object Output Stream} to write the vehicle object into a \texttt{Byte Array Output Stream}. After flushing into a packet, all information is contained as a byte array in the packet to be sent. A benefit for doing that is while receiving the packet and retrieving the information, we can simply reverse the above procedure through \texttt{Byte Array Input Stream} and \texttt{Object Input Stream}, so that the dissembled information is a vehicle instance, and the received vehicle can easily track the information and do operations between two vehicles. Moreover, \texttt{HELLO} packets are sent using \textit{broadcast} scheme, while other packets are sent to either one specific or all of the one hop neighbors via point-to-point communication. Note that we also implemented the without-broadcast version, and that could work on a single local machine. Broadcast only works on distributed machines, e.g., tux machines. 

\subsection{Multithreading}
Multiple threads are ran for each of the vehicle for the parallel jobs that needs to be done:
\begin{itemize}
    \item Each vehicle maintains a \textit{broadcast} thread for broadcasting the \texttt{HELLO} messages. In broadcasting scheme, every vehicle could receive every packets, however, as a simulation, many of them should be discarded, if the packets are from the vehicle out of the 100 meters transmission range. Moreover, the loss rate is assumed to be proportional to the distance between the vehicles that communicate. We implemented these functionality in the receiver end to discard packets.
    \item Each vehicle maintains a \textit{sensor updating} thread for updating the vehicle's information for itself (such as GPS, speed, acceleration, etc.), as well as initiating the regular packet every 10 ms, and sending the vehicle information packet to all its one hop neighbors.
    \item Each vehicle keeps a server socket and this thread listens for all the packets it could receives. Each type of packets has a different processing scheme as in separate methods. 
      \begin{itemize}
          \item All received \texttt{HELLO} messages need to be processed since they do not have sequence numbers, and upon the neighborhood information of the source, the neighborhood and topology information of the receiver might need to be updated. Moreover, no further retransmission of \texttt{HELLO} messages is required. Upon receiving other packets, sequence numbers will be compared with the corresponding one in the vehicle's cache. 
          \item If it is a new \texttt{normal} or \texttt{TC} packet, the source vehicle information then might be utilized by the receiver, and the packet might be further forwarded if such a receiver is a selected MPR. 
          \item And other types of special packets will be utilized or forwarded depending on the destination information in the packet as well as the MPR selector table and \textit{routing table} in the receiver. \textit{Bread First Search} is used for the calculation of the next hop. 
      \end{itemize}
    \item Upon receiving each packet (after discarding those due to out of transmission range and because of loss rate), it may be forwarded. Two different approaches are implemented. One is start a specific thread to forward this particular packet to all its neighbors, and soon close this thread, upon every arrival of packet that should be forwarded. The other is the receiving thread adds every packet that should be forwarded into a queue, and a separate thread specializes in forwarding packets from the queue in FIFO manner. Note that before forwarding a packet, the header should be modified, so that the previous hop is changed to this vehicle's address (or, simply the node ID, in our implementation).
    \item Each vehicle has a separate thread to do the I/O operations with the configuration file. Since the I/O operations are so expensive, we tried to minimize its work. We used concurrency locks for preventing the conflict. The read operation then could be done frequently, as we set to every 50 ms. However, we must reduce the frequency of the write operation on the configuration file, and we set it to every 5 seconds. 
    \item Each following vehicle would have a keyboard listening thread to catch the \texttt{JOIN} and \texttt{LEAVE} request, so that our request could be made by the user at any time, so that our simulation is just like in the real world circumstances. The thread then pass the variable values changed by the event handler to the process, and the procedure routines start.
\end{itemize}

\subsection{Simulation Settings for Vehicle Behavior}
In the simulation application, we assume the lead truck is 10 meters long, and each following vehicle is either 5 or 10 meters long (5m if it is a car, 10m if it is a truck). 

The program would run with one lead truck and up to 10 following vehicles. The wireless nodes are distributed in any area large enough (about 10,000 meters of highway) for the following vehicles and lead truck to travel over a 5 minutes simulation period. All the following vehicles and the lead truck travel in one direction in a highway, e.g. towards East, and that the highway consists of
only two lanes in the same direction. The width of each lane is assumed to be 5 meters. The x-coordinate of the highway is from 0 to 10,000 meters. When the simulation starts, the lead truck is on the right lane with x-coordinate of 10 meter and all the (intended) following vehicles are on the left lane from x-coordinates of 0 to 300 meters in the 10,000 meter length of the highway. At no time can the following vehicles and the lead truck overlap each other. At the beginning of the 5 minutes simulation period, all the following vehicles will use the automated program to form a road train behind the lead truck, and then maintain the road train formation with a pre-specified spacing, i.e. 10-20 meters between the lead truck and the first following vehicle and between any two following vehicle. They will maintain the road train by tracking the position, speed, acceleration and brake control of the lead truck. 

The leading truck would try to keep move smoothly at the speed of 30 m/s, but with a small perturbation of accelerating between $-1$ m/s$^2$ to $1$ m/s$^2$. And it would always be the head of the road train. 

\subsubsection*{Reduced Speed Zone in Simulation}
At the x-coordinate of 4000 meters of the highway, all the vehicles reduce their speed to 20 meter/sec due to a reduced speed zone. All the following vehicles will receive the change of speed from the sensor data packet broadcasted by the lead vehicle and reduce their own speed to 20 meter/sec.

At the x-coordinate of 5000 meters of the highway, all the vehicles would resume their speed to 30 meter/sec (average) since they have passed reduced speed zone. All the following vehicles will receive the change of speed from the sensor data packet broadcasted by the lead vehicle and increase their own speed to 30 meter/sec (average).

At the final phase before the end of the 5 minutes simulation period, all the following vehicles
must leave the road train, travel independently of the lead truck and move to the left lane.

\subsection{Start the Simulation}
A \texttt{main} method in the \texttt{Simulation} class starts the simulation. The user needs to specify the type of the vehicle to be either ``leading'' or ''follow'', and initial x-coordinate and initial speed arguments are optional. A leading truck should always be instantiated before any following vehicle.

\section{Experimental Results}
The experiments are conducted on the distributed \texttt{tux} computers at the university, with the number of vehicles in the VANET to be 2, 4, 6, 8, and 10, respectively. 

The results are reported based on the following 3 performance metrics:
\begin{itemize}
    \item \textbf{End-to-End Latency:} The delay in delivering a packet to the destination which is inclusive of all kinds of delay, i.e. sum of arrival time minus sum of send time divided by total number of connections. 
    \item \textbf{Average Throughput:} This is the measure of the rate at which data being successfully delivered over a communication channel, and is usually measured as the total size of received packets with the unit time. 
    \item \textbf{Packet Loss Rate:} The ratio of data packets lost to those generated by the sources, i.e. (number of packets received) / (number of packets sent). Thus, in the results, the sequence number of packets were sent is traced, and the number of received packets is counted, finally their ratio $PDR$(Packet Delivery Ratio) is calculated, and $(1-PDR)$ is the packet loss rate. 
\end{itemize}


\subsection{Latency}
One critical criteria for VANET is the end-to-end latency, which should be acceptable so that vehicles can timely receive information and do corresponding actions. For instance, if a car ahead breaks unexpectedly and  make an emergency stop needs to be made, every fraction of a second counts. Therefore, we want to see what the latency is under different scenarios. Unlike in previous two projects, using MPRs, the source vehicle which initiates a packet will never receive this packet again, since only MPRs retransmit to their one-hop neighbors except the sender. In this case, we intentionally let the follow vehicles retransmit some of the normal packets back to the leading truck if the packets are directly from the leading truck (i.e., both the source and previous hop of the received packet are the leading truck). In this way, the leading truck could track the round trip time of these packets, and calculate it as the average latency.

\begin{figure}[h!]
    \centering
    \includegraphics[width=\textwidth]{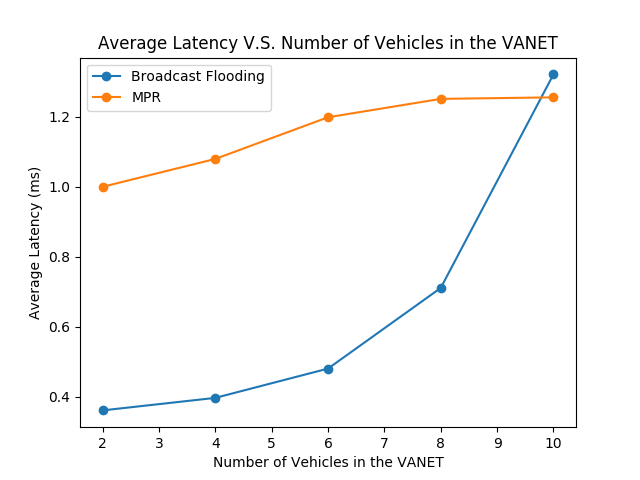}
    \caption{Plot of Average Latency (in milliseconds)}
    \label{fig:latency}
\end{figure}

In Figure 4, we compared the latency by using MPR for retransmission we implemented in this project and using flooding we implemented in the previous project. We plotted the latency time with different number of vehicles in the road train.
Intuitively, as the number of vehicles in the road train grows, the number of packets flooded in the network should grow, so that every node should handle more packets. As a result, latency should grow. As we can see in our figure, with the growing number of vehicles, the average latency in this network of both MPR and flooding grows. This result conforms our expectation.
Remark that in our plot, the MPR latency is higher than the flooding in general. However, we would argue that the reason is that we implemented the flooding retransmission as broadcast, and used point-to-point socket connection for MPR retransmission. The broadcast thread would always be there, but every time we wish to generate a point-to-point connection, we need to establish a new thread and close it after packet is sent. In this way, we would say that the higher latency is due to opening new threads, it does not mean MPR retransmission approach performs worse. Moreover, we can see that the latency grows exponentially for flooding as the number of vehicles increases, while the latency grows sub-linearlly for the MRP retransmission. We would expect MPR retransmission performs better than flooding as more numbers in a VANET.

\subsection{Throughput}

\begin{figure}[h!]
    \centering
    \includegraphics[width=\textwidth]{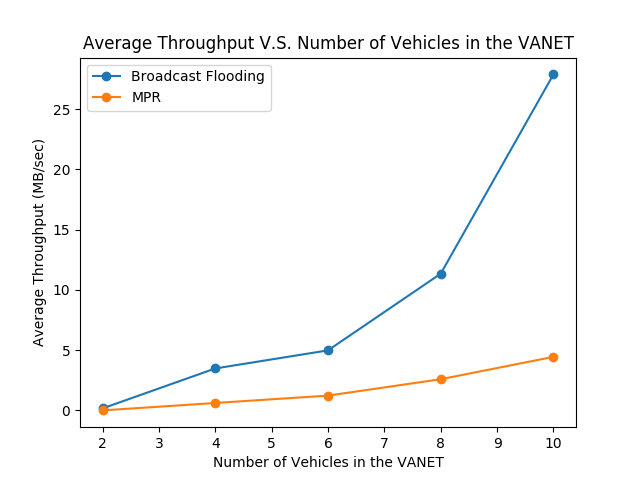}
    \caption{Plot of Average Throughput (MB per second)}
    \label{fig:thruput}
\end{figure}

As shown in Figure \ref{fig:thruput}, with growing number of vehicles in the road train, more packets will be sent and retransmitted, and the throughput in the network should grow. The figure we plot conforms to the trend that we expected. Moreover, the exponential growth pattern for the flooding can be more clearly seen in Figure \ref{fig:thruput}, while using MPR to re-transmit, the throughput is much smaller, even with frequent extra \texttt{HELLO} messages sent to neighbors periodically. Flooding consumes too much traffic for possible unnecessary retransmission, and using MPRs in VANET is much more efficient.

\subsection{Packet Loss Rate}
As with any wireless system, packet loss is always a constant concern. In our experiments we found that packets loss also increased with growing number of vehicles in the road train. The result is plotted in Figure 6. The two approaches have similar growth rate of the average loss rate as the number of vehicles increases. That illustrates that the loss rate is mainly due to the distance-based loss function we deployed in the settings, and has little concern with the network traffic. The stable difference between the two appoaches might because that the neighborhood information of using MPR through \texttt{HELLO} messages is more accurate than flooding. 

\begin{figure}[h!]
    \centering
    \includegraphics[width=\textwidth]{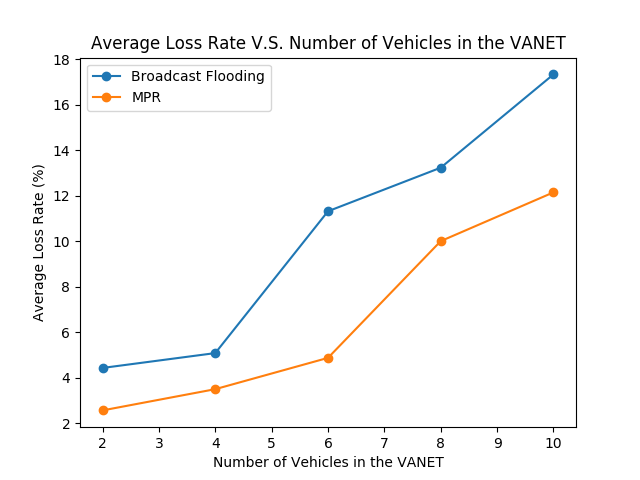}
    \caption{Plot of Average Packet Loss Rate}
    \label{fig:loss}
\end{figure}

\section{Conclusions}
In this paper, we implemented a VANET with a leading truck and multiple following vehicles, and simulated the transmission with different distances between the vehicles. We improved the retransmission by choosing relay points, and that alleviate the flooding problem.
We measured the performance with different number of vehicles in the network, and compared with the results we obtained using broadcast flooding. We conclude that with MPR, the communication in the VANET to form a road train is more efficient and reliable.

\bibliography{references}
\bibliographystyle{plain}





\end{document}